\documentclass[times,twocolumn,final]{elsarticle}

\usepackage{medima}
\usepackage{framed,multirow}
\usepackage{amssymb}
\usepackage{latexsym}
\usepackage{enumitem}
\usepackage{url}
\usepackage{xcolor}
\usepackage{amsmath,amssymb,amsfonts}
\usepackage{algorithmic}
\usepackage{graphicx}
\usepackage{textcomp}
\usepackage{booktabs}
\usepackage{multirow}
\usepackage{soul}
\usepackage{color}

\definecolor{newcolor}{rgb}{.8,.349,.1}

\journal{Preprint}

\begin{document}

\verso{Abdullah F. Al-Battal \textit{et~al.}}

\begin{frontmatter}

\title{Multi-target and multi-stage liver lesion segmentation and detection in multi-phase computed tomography scans}%

\author[1,2]{Abdullah F. \snm{Al-Battal}\corref{cor1}}

\cortext[cor1]{Corresponding author: 
  email: aalbatta@ucsd.edu}

\author[3,4]{Soan T. M.  \snm{Duong}}

\author[3]{Van Ha  \snm{Tang}}

\author[3]{Quang Duc  \snm{Tran}}

\author[3]{Steven Q. H. \snm{Truong}}

\author[5]{Chien \snm{Phan}}

\author[1]{Truong Q.  \snm{Nguyen}}

\author[1]{Cheolhong \snm{An}}

\address[1]{Electrical and Computer Engineering Department,
        UC San Diego, La Jolla, California, USA}
\address[2]{Electrical Engineering Department, King Fahd University of Petroleum and Minerals, Saudi Arabia}
\address[3]{Vinbrain JSC, Hanoi, Vietnam}
\address[4]{Department of Computer Science, Le Quy Don Technical University, Hanoi, Vietnam}
\address[5]{Diagnostic Imaging Department, University Medical Center at Ho Chi Minh City, Vietnam}

\begin{abstract}
%%%
\small
Multi-phase computed tomography (CT) scans use contrast agents to highlight different anatomical structures within the body to improve the probability of identifying and detecting anatomical structures of interest and abnormalities such as liver lesions. Yet, detecting these lesions remains a challenging task as these lesions vary significantly in their size, shape, texture, and contrast with respect to surrounding tissue. Therefore, radiologists need to have an extensive experience to be able to identify and detect these lesions. Segmentation-based neural networks can assist radiologists with this task. Current state-of-the-art lesion segmentation networks use the encoder-decoder design paradigm based on the UNet architecture where the multi-phase CT scan volume is fed to the network as a multi-channel input. Although this approach utilizes information from all the phases and outperform single-phase segmentation networks, we demonstrate that their performance is not optimal and can be further improved by incorporating the learning from models trained on each single-phase individually. Our approach comprises three stages. The first stage identifies the regions within the liver where there might be lesions at three different scales (4, 8, and 16 mm). The second stage includes the main segmentation model trained using all the phases as well as a segmentation model trained on each of the phases individually. The third stage uses the multi-phase CT volumes together with the predictions from each of the segmentation models to generate the final segmentation map. To demonstrate its ability to generalize beyond liver lesions, we also test our approach on brain lesions in multi-contrast magnetic resonance imaging (MRI). Overall, our approach improves relative liver lesion segmentation performance by 1.6\% while reducing performance variability across subjects by 8\% when compared to the current state-of-the-art models.
%%%%
\end{abstract}


\end{frontmatter}

%\linenumbers

%% main text
%%%%%%%%%%%%%%%%%%%%%%%%%%%%%%%%%%%%%%%%%%%%%%%%%%%
%%%%%%%%%%%%%%%%%%%%%%%%%%%%%%%%%%%%%%%%%%%%%%%%%%%
\section{Introduction}
\label{sec:introduction}
Liver cancer is the sixth most commonly occurring cancer and the third most common cause of cancer death worldwide \citep{sung2021global}. It is also the second most common cause of premature death from cancer \citep{rumgay2022global}. Detecting and identifying lesions within the liver is a crucial step in the diagnosis and treatment of liver cancer. Computed tomography (CT) scans play a pivotal role in the detection and segmentation of lesions and tumors in the liver, serving as a fundamental technique in both diagnostic assessments, and the planning and tracking of treatment procedures \citep{elbanna2021computed}. The accurate identification and delineation of hepatic abnormalities are crucial for effective treatment planning, yet they present significant challenges due to the complexity and variability of the liver anatomy across subjects, the variable appearance of lesions including their shape, size, and contrast with surrounding healthy tissue, as well as imaging artifacts and noise. As a result, radiologists, despite extensive training and experience, are still challenged by this task where the recall of liver lesions can be as low as 72\% for lesions of sizes in the range of 10-20 mm and 16\% for lesions smaller than 10 mm \citep{wiering2007comparison,freitas2021imaging}. Consequently, there is a need for computerized approaches to assist radiologists in detecting and segmenting lesions within the liver.

Multi-phase CT scans are used by clinicians to enhance liver lesion diagnostic accuracy through contrast enhancement in different phases, namely the delay, arterial, and venous phases \citep{oliva2004liver,bae2010intravenous}. During the arterial phase, the CT scan captures images shortly after contrast injection, highlighting the vascular nature of certain lesions, such as hepatocellular carcinoma, by showing their hyperenhancement due to their arterial blood supply. The venous phase follows, where contrast washout in lesions and enhancement of the surrounding liver parenchyma can help to distinguish between benign and malignant lesions, as many malignancies show a distinct contrast behavior compared to the liver. The delay phase, taken several minutes after contrast administration, provides further insight by showing the retention of contrast in certain lesions, such as hemangiomas and fibrotic changes, aiding in their identification and differentiation from other pathologies. This multi-phase approach allows for an overall improved evaluation of liver lesions by leveraging the dynamics of contrast uptake and washout, which increases lesion saliency with respect to healthy tissue; increasing detection and segmentation sensitivity \citep{oliva2004liver}. 

Segmentation algorithms aim to identify, localize, and provide a dense label for each pixel that belongs to the objects of interest within an image . This also extends to 3D medical image volumes such as those encountered in CT and MRI scans where a dense label is assigned for all voxels within the objects of interest \citep{lakare20003d}. Current state-of-the-art segmentation approaches rely on deep learning models. Prior to deep learning, many traditional segmentation methods achieved relative success in isolating objects of interest from its surrounding for segmentation purposes \citep{bulu2007comparison}. These traditional methods, however, often relied on handcrafted features to identify objects of interest, which limited their ability to generalize beyond their specified application. In medical imaging, methods such as intensity level thresholding \citep{liao2001fast}, region growing \citep{lin2000unseeded}, active contours, block matching, shape fitting \citep{dickens2002volumetric} and watershed transformation \citep{salman2010segmentation} were relatively successful at localized segmentation and object isolation from its surrounding. They were, however, less successful at global object localization and segmentation within the context of a whole image or a 3D medical scan.

In contrast to traditional methodologies that depend on pre-defined features, deep learning strategies employ trainable parameters. These parameters are dynamically adjusted throughout training to enhance performance on the designated task, thereby improving their ability to generalize \citep{rumelhart1986learning}. This learning approach is key to improve their generalization capabilities and allows deep learning methods to excel in global object localization and segmentation. In the context of segmentation, deep learning models use an encoder to extract features from the input image, and a decoder to spatially contextualize these features \citep{long2015fully}, generating a pixel-wise segmentation map for 2D images or a voxel-wise map for 3D images. Within the domain of medical image segmentation, convolutional neural networks (CNNs) stand out as the best performing models \citep{u_net_2015,zhou2019unet++,isensee2021nnu,oktay2018attention}. CNNs utilize convolutional kernels within the encoder to extract features and in the decoder to spatially contextualize these features. Besides CNNs, transformers have also shown significant promise in medical image segmentation by processing images on a patch-wise basis in the encoder instead of relying on convolutional kernels \citep{hatamizadeh2022unetr,hatamizadeh2021swin}. Both CNNs and transformers have their own advantages and disadvantages. CNNs excel at capturing spatial hierarchies and local features within images, making them particularly suited for detailed and texture-rich medical images, but they struggle with long-range dependencies. On the other hand, transformers handle global context well while less capable of capturing local features without excessively large datasets and higher computational resources \citep{dosovitskiy2020image,liu2021swin}.

Many deep learning methods aimed to solve the difficult problem of liver lesion segmentation. These models perform better when trained on 3D image volumes rather than 2D image slices or 2.5D multi-slice images \citep{bilic2023liver}. Fully convolutional networks (FCN) based on the UNet architecture are the most successful at this problem so far \citep{u_net_2015,bilic2023liver}. The UNet architecture uses a symmetric structure with a contracting path (the encoder) to capture context and extract features and an expanding path to localize features (the decoder) with skip connections between the encoder and decoder to combine features at different stages of the network. Several developments have been proposed to the UNet architecture to improve its segmentation and efficiency performance. These developments targeted different components of the network. Notably, these improvements included modifications to the convolutional block, such as the integration of residual or bottleneck blocks, which aid in mitigating the vanishing gradient problem and improving feature representation without substantially increasing computational complexity \citep{he2016deep,khanna2020deep}. Additionally, advancements have been made in the architecture's skip connections, with the introduction of nested and multi-stage connections that facilitate the model's ability to capture and integrate multi-scale contextual information more effectively \citep{zhou2019unet++}. Moreover, the incorporation of attention mechanisms within the skip connections further refines the model's focus on relevant features, significantly enhancing segmentation precision by selectively emphasizing important spatial features while suppressing less relevant information \citep{oktay2018attention}.

\begin{figure*}[ht]
\vspace{0.05cm}
\begin{minipage}[b]{.33\linewidth}
  \centering
  \centerline{\includegraphics[height=3.4cm, width=1\linewidth]{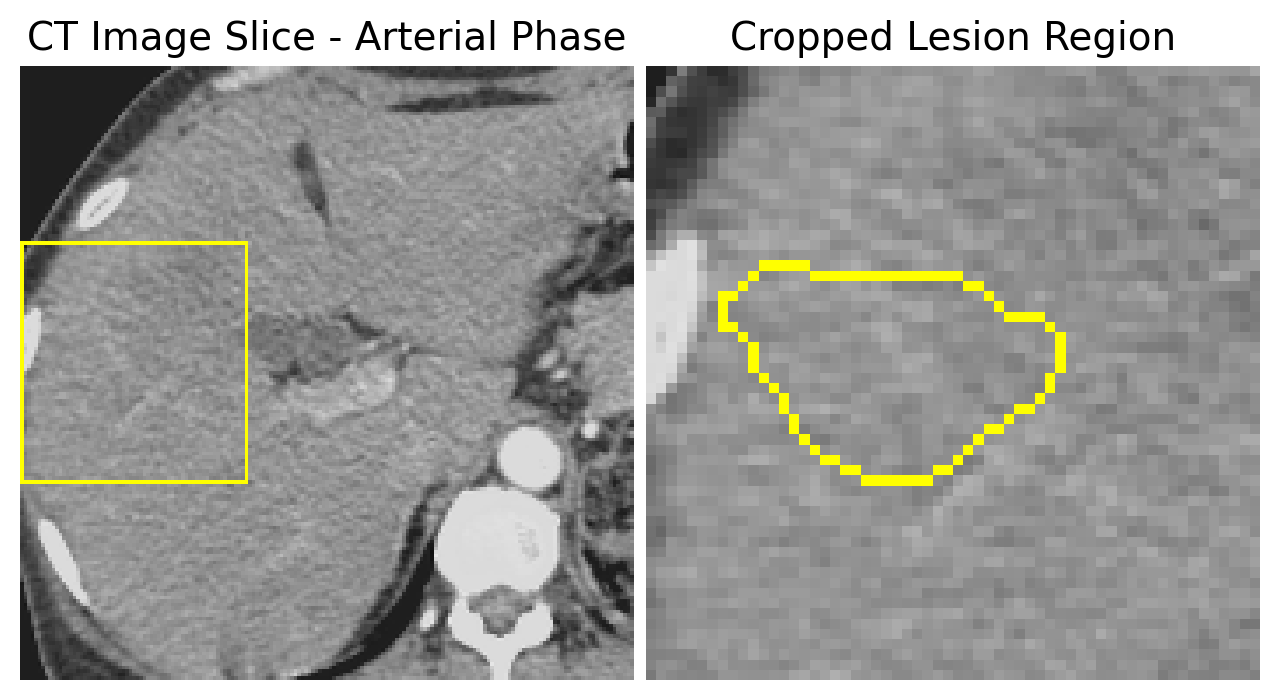}}
  \vspace{-0.05cm}
  \centerline{\footnotesize(a)}
\end{minipage}
\begin{minipage}[b]{.33\linewidth}
  \centering
  \centerline{\includegraphics[height=3.4cm, width=1\linewidth, trim = {0cm 0cm 0cm 0cm}, clip]{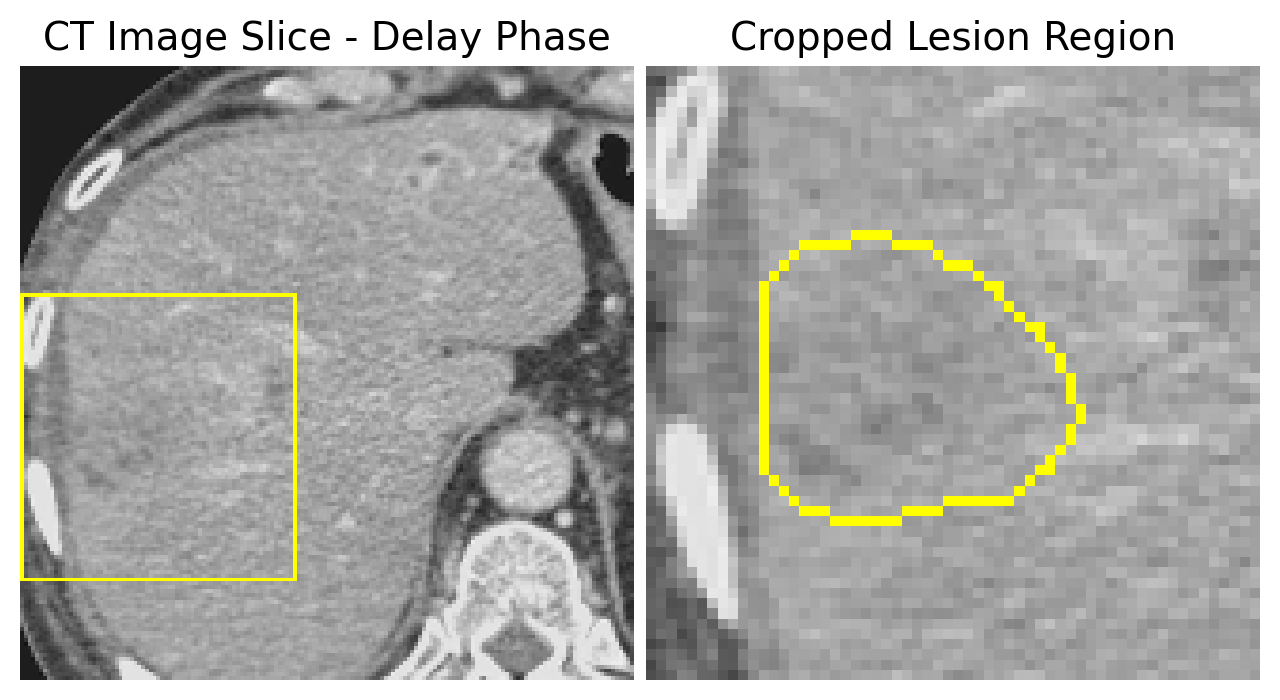}}
    \vspace{-0.05cm}
  \centerline{\footnotesize(b)}
\end{minipage}
\begin{minipage}[b]{.33\linewidth}
  \centering
  \centerline{\includegraphics[height=3.4cm, width=1\linewidth, trim = {0cm 0cm 0cm 0cm}, clip]{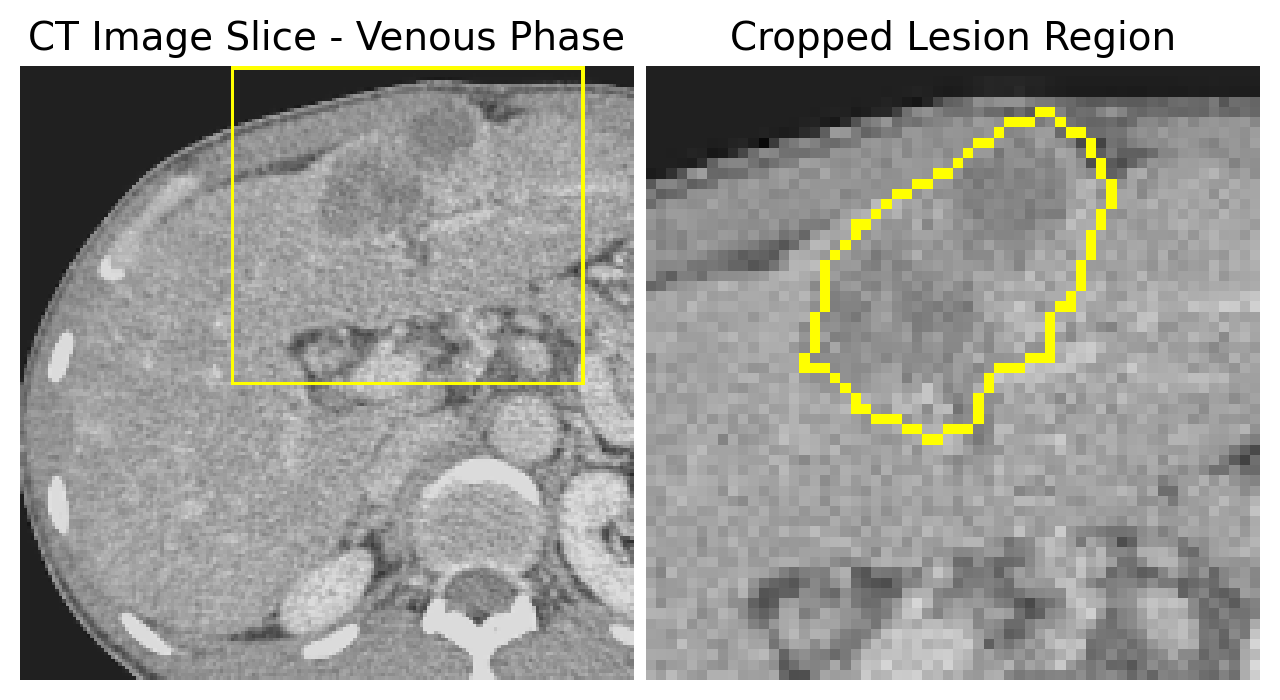}}
    \vspace{-0.05cm}
  \centerline{\footnotesize(c)}
\end{minipage}
\\
\vspace{-0.7cm}
\caption{Three slices from different subjects in the liver lesion dataset (a)-(c). Each slice is acquired at a different phase and cropped to the liver region. The lesion region of interest is highlighted in a yellow bounding box. Within each of the cropped regions, the boundary of the lesion is outlined in yellow.}
\label{fig:example_slices}
\vspace{-0.2cm}
\end{figure*}

In multi-phase and multi-contrast segmentation, state-of-the-art models like MedNext \citep{roy2023mednext} and nnUNet \citep{isensee2021nnu} leverage the distinct phases of contrast-enhanced CT and MRI images as separate channel inputs to utilize the unique information each phase and contrast provide to enhance segmentation accuracy. These models have demonstrated remarkable proficiency in delineating objects of interest by capturing the diverse dynamics of objects appearance across different contrast phases, which is crucial for accurate lesion characterization. However, it should be noted that current leading methods do not incorporate attention or feature fusion mechanisms within their skip connections, which has the potential to refine feature selection and improve segmentation precision by focusing on areas of interest within the image. Contrary to these state-of-the-art approaches, our approach extends the multi-phase segmentation strategy by first employing individual models to extract features from each phase independently, improving the recovery of distinct information presented in each. Subsequently, another model is utilized to extract features from the CT volumes and the predicted segmentation maps, aiming for a global segmentation enhancement. This is followed by a targeted enhancement process for the segmentation of each lesion individually. The proposed approach enhances the utilization of both global context and lesion-specific details from each of the phases, improving the accuracy, consistency and granularity of liver lesion segmentation.

%%%%%%%%%%%%%%%%%%%%%%%%%%%%%%%%%%%%%%%%%%%%%%%%%%%
\subsection{Clinical Relevance and Contribution}
\label{sub-sec:clin_sig}

Accurate and consistent detection, localization and segmentation of liver lesions in CT scans is a key step in diagnosing and managing liver cancer. This precision is crucial not only for identifying the presence of malignant tissues but also for accurately assessing their size, shape, and location within the liver. Such detailed information is necessary for oncologists and surgeons in formulating tailored treatment plans, including the selection of appropriate surgical interventions and targeted therapies as well as monitoring disease progression. Otherwise, the consequences can be severe, potentially leading to the progression of cancer and premature fatalities if lesions are missed or under-characterized. On the other hand, overestimation of lesion size or incorrect localization can lead to unnecessary or overly aggressive treatments, subjecting patients to risks and complications. Such errors may result in surgeries that remove more healthy tissue than necessary or in the selection of treatment modalities that do not optimally target the cancer, adversely affecting the patient's quality of life and survival chances.

In this paper we propose a multi-stage multi-target segmentation framework that is based on fully convolutional neural networks. The framework comprises three stages. The first stage identifies the regions within the liver where there might be lesions at three different scales (4, 8, and 16 mm). The second stage includes the main segmentation model trained using all the phases as well as a segmentation model trained on each of the phases individually. The third stage uses the multi-phase CT volumes together with the predictions from each of the segmentation models to generate the final segmentation map. Contrary to current state-of-the-art multi-phase segmentation approaches that incorporate different phases as different input channels to the model, our approach incorporates learning from each individual phase. In the main segmentation model we design a feature fusion and attention (FF\&A) module that improves the ability of the model to combine features from the decoder and encoder in the skip connection. The module mixes features from the previous stage in the decoder and features from the encoder in the skip connection and extracts features from the mixture to generate spatial weighting of the features coming from the encoder on a coarse (using projected axial pooling) and a fine basis.

We test the segmentation and detection performance of our approach on a three-phase CT dataset. The dataset contains scans from 354 subjects annotated with liver lesion segmentation labels. Each subject has 3 contrast enhanced scans at three different phases which are the arterial, delay, and venous phases. Example scans from this dataset showing slices from each of the phases are shown in Fig. \ref{fig:example_slices}. In this paper, we compare the proposed framework as well as the main segmentation model of the framework (the second stage of the proposed framework) to the current state-of-the-art segmentation models, which are the MedNext \citep{roy2023mednext}, nnUnet \citep{isensee2021nnu}, SwinUNetR \citep{hatamizadeh2021swin}, and Model Genesis \citep {zhou2021models} models. Both our proposed segmentation model and the overall framework improved the segmentation and detection performance on a global, and by-subject manner; improving the overall segmentation accuracy while reducing performance variability across subjects. Beyond liver lesion, we also tested our approach on the BraTS \citep{menze2014multimodal,bakas2017advancing,bakas2018identifying} Brain Tumor MRI dataset to demonstrate its ability to accurately detect and segment anatomical structures of interest in multi-phase and multi-contrast medical images. Overall, the major contributions of our work are:

\begin{enumerate}[label=\Alph*)]
    \item A multi-stage multi-target segmentation framework for liver lesions in multi-phase CT scans that leverages fully convolutional neural networks; improving liver lesion segmentation and detection.
    \item A fully convolutional stage that identifies possible regions containing lesions at three different scales (4, 8, and 16 mm) to highlight areas of the liver where radiologists and clinicians might need to investigate more thoroughly for possible tumors beyond the final segmentation map.
    \item A segmentation strategy that incorporates features learned from individual phases in both the encoder and decoder in addition to the multi-phase segmentation model for an improved feature extraction and spatial contextualization.
    \item A feature fusion and attention (FF\&A) module in the skip connections of the main segmentation model to integrate and emphasize relevant features from both the encoder and decoder paths for an enhanced segmentation spatial focus.
\end{enumerate}

%%%%%%%%%%%%%%%%%%%%%%%%%%%%%%%%%%%%%%%%%%%%%%%%%%%
%%%%%%%%%%%%%%%%%%%%%%%%%%%%%%%%%%%%%%%%%%%%%%%%%%%
\section{Related Works}
\label{sec:related}

%%%%%%%%%%%%%%%%%%%%%%%%%%%%%%%%%%%%%%%%%%%%%%%%%%%
\subsection{Encoder-Decoder Architectures for Semantic Segmentation}
\label{sub-sec:related_encode}
Deep learning models based on the encoder-decoder architecture with skip connections are the current state-of-the-art for medical image segmentation regardless of the imaging modality or the target anatomical structures \citep{zhou2019unet++, jha2020doubleu, jha2021comprehensive, isensee2021nnu, roy2023mednext, hatamizadeh2021swin}. These architectures, which use a contracting path (the encoder) followed by an expansive path (the decoder), encode spatial information in an image and then contextualizes it spatially to generate a mask map where each pixel, or voxel for 3D images, is represented individually based on the class it belongs to \citep{badrinarayanan2017segnet}. Skip connections which combine features from the layers of the encoder and decoder at the same depth level improve the overall performance of these architectures significantly. These skip-connections transfer the feature map at each level and either concatenate it with or add it to the feature map in the expansive path, forming the well-known UNet architecture \citep{u_net_2015}. The most significant improvements of the UNet architecture either targeted these skip connections \citep{zhou2019unet++, oktay2018attention}, the convolution block within the encoder and decoder \citep{alom2019recurrent,khanna2020deep}, or self-configurability of the network depth and width \citep{isensee2021nnu}. Furthermore, using deep supervision, \cite{zhou2019unet++}, and  \cite{zhu2017deeply} were able to improve the performance of UNet in segmentation tasks for various medical imaging applications.

Among the improvements that focused on redesigning the skip connections of the UNet architecture, UNet++ \citep{zhou2019unet++}, Attention UNet \citep{oktay2018attention}, and MultiResUNet made large strides in improving the overall segmentation capabilities of the UNet architecture across different medical image segmentation tasks. UNet++ uses a nested architecture that refines skip connections using multiple interconnected convolutional networks at different depths, enabling an increased mixture of features across the network. Attention UNet incorporates attention gates within its skip connections, focusing the model spatially on relevant image regions by selectively emphasizing important features while suppressing less relevant information. Other models such as the ResUNet, UNetR \cite{hatamizadeh2022unetr}, SwinUNetR \citep{hatamizadeh2021swin}, and MedNext \citep{roy2023mednext} improved on the UNet architecture by modifying the design of the convolutional block or replacing it with a transformer-based block. The ResUNet architecture replaces the convolutional blocks within the encoder and decoder with residual convolutional blocks while the MedNext model uses a residual bottleneck convolutional block, which is a 3D version of the ConvNext block proposed by \cite{liu2022convnet}. The UNetR and SwinUNetR models replaced the encoder with a transformer-based encoder using the VIT-B model and the Swin Transformer, respectively. The MultiResUNet, on the other hand, included modifications to both the convolutional block as well as the skip connections, by incorporating multi-residual paths inspired by DenseNet in the first and successive residual blocks in the second. Transformers encoders in general, however, struggle with smaller objects and extracting localized dense representations as they encode features using patch-based manner. The Swin Transformer aimed to mitigate this problem with shifted windows, but they still lag behind in their ability to recover small objects such as lesions in their early stages. 

\begin{figure*}[ht]
\vspace{-3mm}
    \centering
    \includegraphics[width = 0.99\linewidth, height=8cm, trim = {2.5cm 0.25cm 0.9cm 0.25cm}, clip]{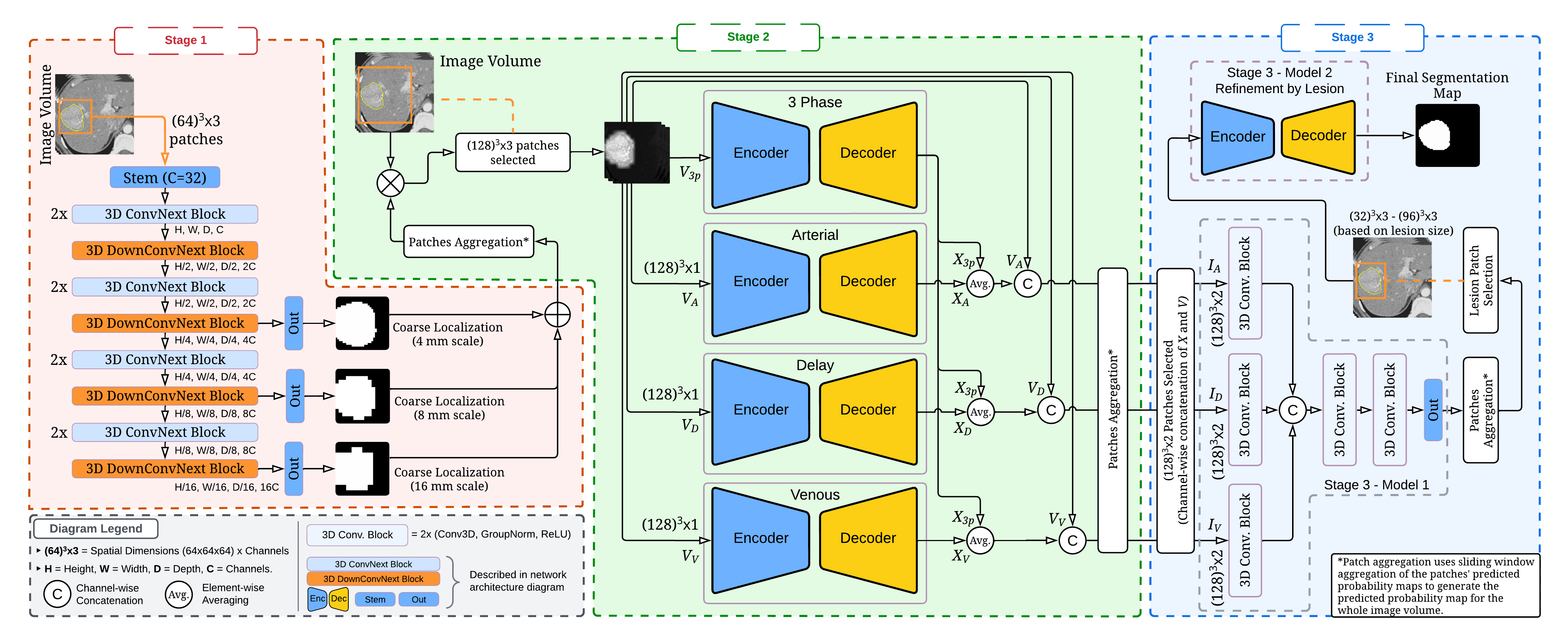}

\vspace{-3mm}
\caption{The proposed framework structure with its different stages. The three outputs of stage 1 are converted to a heatmap that weights the input to stage 2. The outputs of stage 2 are concatenated with the CT image volume from each phase and then fed to stage 3. The structure of the models in stage 2 and model 2 in stage 3, as well as the structure of the Stem, Output, 3D ConvNext, and 3D DownConvNext blocks are outlined in Fig. \ref{fig:model}.}
\vspace{-2mm}
\label{fig:framework}
\end{figure*}

%%%%%%%%%%%%%%%%%%%%%%%%%%%%%%%%%%%%%%%%%%%%%%%%%%%
\subsection{Liver Lesion Segmentation in Single- and Multi-Phase CT Scans}
\label{sub-sec:related_lesion_liver_ct}
Deep learning models based on the UNet architecture are the most widely used and best performing for the task of liver lesion segmentation in single-phase CT scans \citep{bilic2023liver}. These models include the current state-of-the-art model, nnUNet \citep{isensee2021nnu}. The nnUNet model uses the original UNet architecture and incorporates a self-configuration approach that modifies the networks depth, width, and under-sampling stages, among others, based on the dataset footprint in terms of the target anatomical structure intensity and spatial characteristics. The Model Genesis \citep{zhou2021models} UNet model, which uses self-supervised learning as a pre-training approach to learn transferable image representations, also performs comparably to the nnUNet model. Transformer models in general do not perform on bar with CNN models due to the relative small size of lesions to the overall scan. However, the SWinUNetR model can achieve comparable results due to the use of shifted windows which improves local and small feature extraction for dense segmentation predictions. \cite{chen2020octopusnet} proposed OctopusNet, which is a CNN-based architecture that uses separate encoder branches and a single decoder branch for multi-phase medical image segmentation. In their tests, \cite{chen2020octopusnet} found that this approach improves performance when compared to using the different phases as input channels. Nevertheless, current state-of-the-art approaches uses the latter network design.

Although these models were tested on multi-phase and multi-contrast datasets such as the Brain tumor dataset (BraTS), they were not tested on lesion segmentation in multi-phase CT scans prior to the work we present in this paper. At the time of this paper writing, there are currently no publicly available multi-phase liver lesion segmentation datasets. Therefore in our approach we use an internally created three-phase dataset that was annotated and rated by two different radiologists for 354 subjects with primary and secondary liver tumors. Each subject was scanned at the arterial, delay, venous phases of contrast injection. Multi-phase and multi-contrast models use each of the phases or contrast images as a separate input channel to the overall model while maintaining the same structure of the model overall. For example, a model trained on segmenting a three-phase CT scan will have a single input of three channels, each channel representing each of the phases individually. Although this approach incorporates information from all of the phases, we demonstrate in our work that this is not optimal and that incorporating models trained on each of the phases individually enhances the segmentation performance significantly.

%%%%%%%%%%%%%%%%%%%%%%%%%%%%%%%%%%%%%%%%%%%%%%%%%%%
%%%%%%%%%%%%%%%%%%%%%%%%%%%%%%%%%%%%%%%%%%%%%%%%%%%
\section{Proposed Method}
\label{sec:method}
Effective identification and segmentation of lesions in the liver benefits significantly from imaging at different phases post contrast injection as the response of these tumors to the contrast agent at different times allow them to be more distinguishable from surrounding tissue.  Hence, it is important to be able to extract features from these scans in a matter that allows the segmentation model to segment them accurately. Therefore, we design our framework, which is composed of three stages as shown in Fig. \ref{fig:framework}, to incorporate a segmentation model trained using all the phases as multi-channel input as well as a segmentation model trained on each of the phases individually. The outcomes of these models together with the CT volumes are fed into the third stage, which is a segmentation correction and refinement stage, to generate the final segmentation map. Prior to these two stages we feed the CT volume to the first stage, which identifies the areas within the liver where there might be lesions at three different scales (4, 8, and 16 mm). In the main segmentation model we design a feature fusion and attention (FF\&A) module that improves the ability of the model to combine features from the decoder and encoder in the skip connections. In this section, we explain our approach in detail, starting with the augmentation and pre-processing techniques that are used during training to promote generalization and robustness.

\begin{figure*}[ht]
\vspace{-1mm}
    \centering
    \includegraphics[width = 0.99\linewidth, height=8cm, trim = {0.9cm 1cm 0.25cm 0.25cm}, clip]{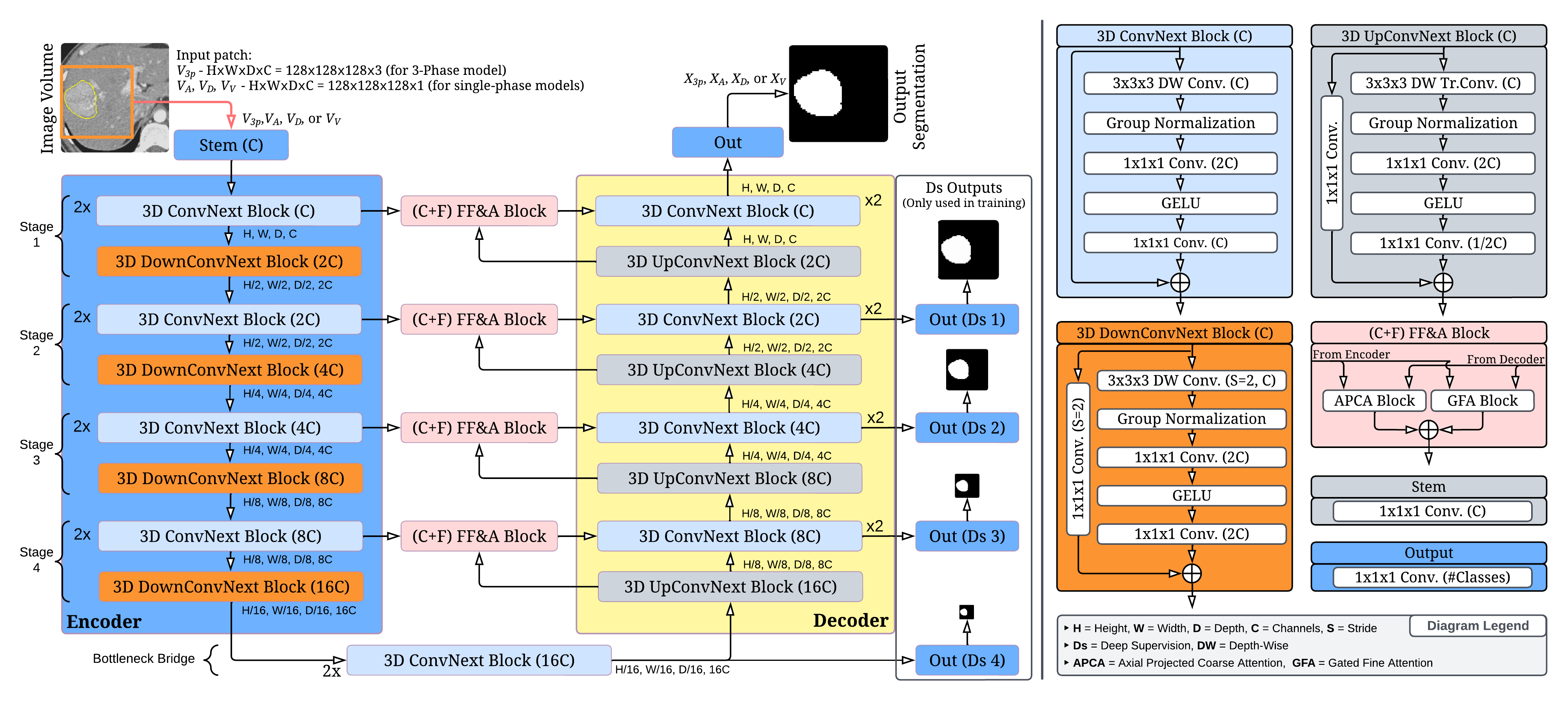}

\begin{minipage}[b]{.57\linewidth}
  \centering
  \centerline{\hspace{0.2cm}\footnotesize(a)}
\end{minipage}
\begin{minipage}[b]{.4\linewidth}
  \centering
  \centerline{\hspace{1.2cm}\footnotesize(b)}
\end{minipage}

\vspace{-1mm}
\caption{The architecture and structure of the proposed main segmentation model in stage 2 of the framework (a) with the different building blocks outlined (b): The convolutional block in the encoder and decoder branches, the upsampling  convolutional block, the attention and feature fusion block, and the stem and output blocks. The spatial dimension and number of feature channels at the output of each of these blocks are outlined in (a). The structure of model 2 in stage 3 of the framework is the same as the main model in (a), but uses only 3 stages (stages 1, 2, and 3) instead of 4 in the encoder and decoder together with the bottleneck bridge.}
\vspace{-2mm}
\label{fig:model}
\end{figure*}

%%%%%%%%%%%%%%%%%%%%%%%%%%%%%%%%%%%%%%%%%%%%%%%%%%%
\subsection{Pre-Processing and Augmentation}
\label{sub-sec:method_pre}
Pre-processing and augmentation while training neural networks is an essential step to promote robustness and generalization. We use them in our proposed approach to induce variabilities into the training set stemming from the variabilities inherent in scans that would be present when the model is deployed. The proposed augmentation methods can be categorized into three main categories: a) intensity-based to account for different imaging devices and b) geometrically rigid and c) elastic transformations to account for variations in anatomical structures' shape, size and elasticity. Prior to augmentation, we pre-process the CT volume images and prepare them for our segmentation framework by clipping and normalizing their intensity values and randomly selecting a patch of size $128\times128\times128$ voxels, which are the dimensions of the model input image. All the CT volumes are resampled to an isotropic spatial spacing of $1\times1\times1$ mm.

For intensity-based augmentation, we randomly choose a set of options to change the visual properties of CT scans during training. We adjust properties like brightness, contrast, and noise levels.  These changes teach the model to identify important features even when scans are created using different CT devices or reconstruction settings. Additionally, we simulate how a patient might be positioned slightly differently for each scan. We do this by rotating, shifting, cropping, resizing, and flipping the images. This helps the model become more reliable, ensuring it can accurately analyze scans even when there are variations in how the patient was situated during the imaging process. Additionally, we use affine transformations that include shearing and scaling to mimic the way soft tissues might stretch or compress. We also introduce elastic deformations that simulate the natural variations in shape and position of internal organs. These variations can be caused by factors like breathing, differences in patient body size, or even the presence of tumors or other abnormalities. By simulating these tissue variabilities, we make the model more robust in recognizing important anatomical features despite the inevitable differences between individual patients.

 %%%%%%%%%%%%%%%%%%%%%%%%%%%%%%%%%%%%%%%%%%%%%%%%%%%
\subsection{The Liver Lesion Segmentation Framework}
\label{sub-sec:method_frame_work}

%%%%%%%%%%%%%%%%%%%%%%%%%%%%%%%%%%%%%%%%%%%%%%%%%%%
\subsubsection{Stage 1: Lesion localization and Patch Flagging Model}
\label{sub-sub-sec:model_stage_1}
This stage is tasked with identifying the areas within the liver that might contain lesions at three different scales (4, 8, and 16 mm). This stage's model is outlined in Fig. \ref{fig:framework}. The first component of the model is a stem that expands the input patch feature width from 3 (a channel for each phase) to 32 channels. The model uses five convolutional blocks. Each convolutional block contain two 3D ConvNext blocks followed by a 3D down sampling ConvNext block, apart from the last convolutional block, which does not use a down sampling block. The outputs from the third (prior to down sampling), fourth (also, prior to down sampling), and fifth convolutional blocks are then fed to an output convolutional layer to generate a segmentation map at each of the scales, namely 4, 8, and 16 mm. These segmentation maps work as an area or patch flagging mechanism where patches at each of the scales are either flagged (containing lesion) or not. The patches are then combined to form a compound heatmap that highlight areas within the liver where the model believes there are lesions.

%%%%%%%%%%%%%%%%%%%%%%%%%%%%%%%%%%%%%%%%%%%%%%%%%%%
\subsubsection{Stage 2: Lesion Segmentation Model}
\label{sub-sub-sec:model_stage_2}
In this stage we train the encoder-decoder segmentation models outlined in Fig. \ref{fig:model}. We train a main model that takes all the phases as input channels as well as a model for each of the phases individually. The first component of this model is also a stem that expands the input patch feature width from 3 (a channel for each phase) to 32 channels. The model uses four convolutional blocks in the encoder and four convolutional blocks in the decoder with a bridge bottleneck after the last encoder block to connect the encoder with decoder. Each convolutional block contains two 3D ConvNext blocks followed by a 3D down sampling ConvNext block while the bridge bottleneck contains only two 3D ConvNext blocks. The structure of the ConvNext and down sampling ConvNext blocks are outlined in Fig. \ref{fig:model} (b). ConvNeXt blocks improve upon prior convolutional blocks design by offering greater efficiency through depth-wise separable convolutions, and incorporating transformer-inspired design paradigms such as the inverted bottleneck design. This enables ConvNext-based models to perform on bar with transformer-based models on coarse computer vision tasks such as classification while outperforming them on dense prediction tasks such as segmentation, especially for small objects. To train the model, we incorporate deep supervision where the output of each convolutional block in the decoder at each depth level (N) is fed to an output linear projection layer that contracts the channel space from $2^{(N-1)}  C$ to $2$, which represents the number of classes (background versus lesion). 

\begin{figure}[h]
\includegraphics[width = 0.99\linewidth, height=4.55cm ]{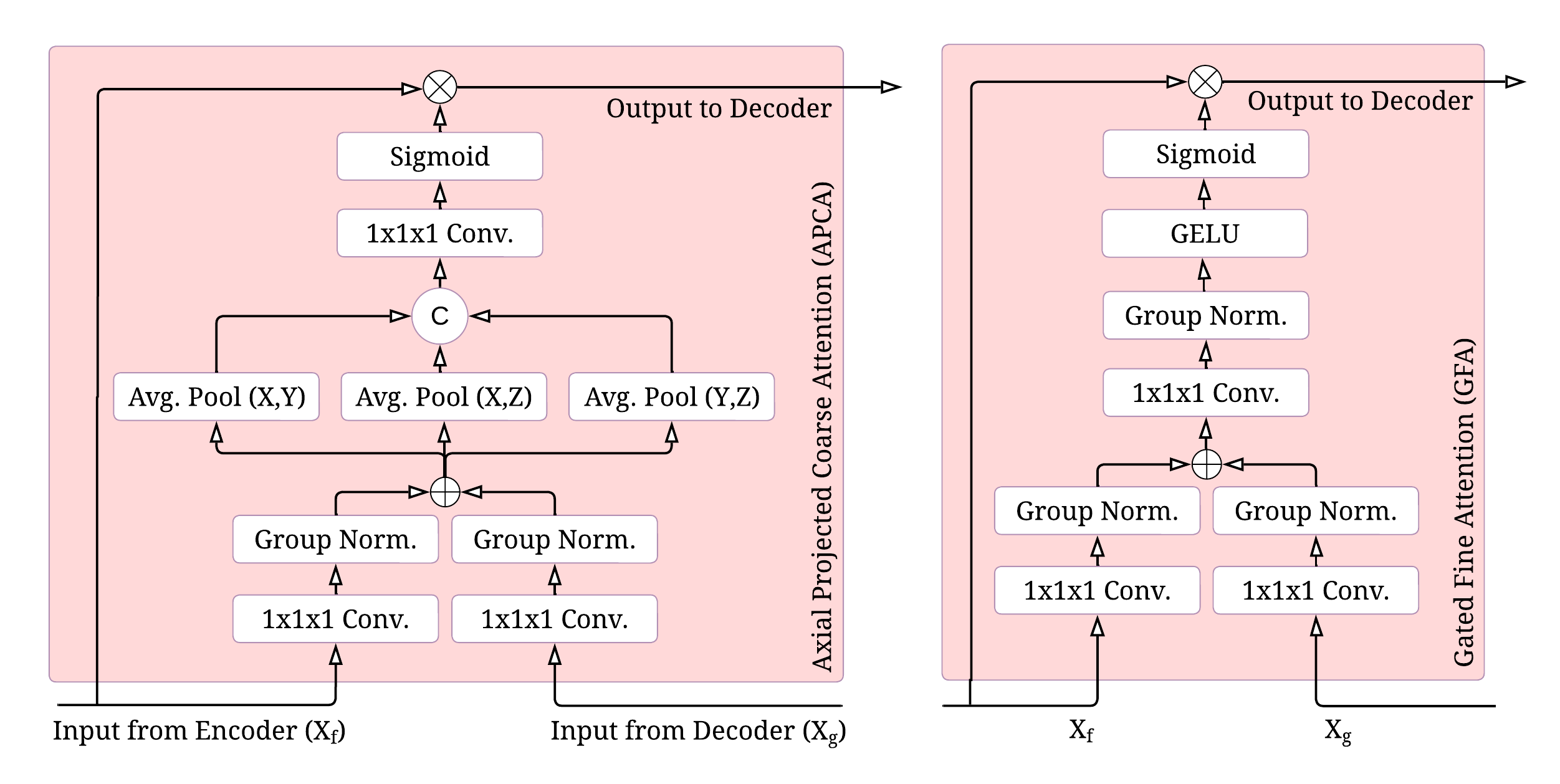}
\vspace{-4mm}
\caption{ The structure of the Axial Projected Coarse Attention (APCA) module and the Gated Fine Attention module (GFA), which are the two components of the Coarse+Fine Feature Fusion \& Attention Module.}
% \vspace{-6mm}
\label{fig:att_blocks}
\end{figure}

To improve feature mixing in the skip connection between the encoder and decoder, we incorporate a Coarse+Fine Feature Fusion \& Attention Module (C+F FFA). This module includes two feature fusion and attention mechanisms; the first is Axial Projected Coarse Attention (APCA) while the second is Gated Fine Attention (GFA), to refine feature maps by emphasizing spatially relevant information. The APCA mechanism processes features from the encoder ($X_f$) and decoder ($X_g$) layers, using 1$\times$1$\times$1 convolutions and group normalization to create a compact representations in the feature space. These are mixed and then projected across each of the three spatial dimensions using adaptive average pooling, concatenation, and further feature extraction using convolutional layers to generate coarse and smooth attention maps that is less susceptible to noise due to axial based projection and averaging. These maps are used to spatially weight the feature map from the encoder ($X_f$), to enhance relevant features. The GFA module, in parallel, processes the same feature sets with 1$\times$1$\times$1 convolutions and group normalization, combining them without axial projection to maintain fine details of representations within the feature map. A subsequent convolution produces a spatial attention map, modulating ($X_f$) to focus on important spatial regions. Together, these mechanisms enable focusing on salient features of interest, which enhances the segmentation accuracy by acknowledging spatial relationships within the image. The detailed structure of both modules is outlined in Fig. \ref{fig:att_blocks}.

%%%%%%%%%%%%%%%%%%%%%%%%%%%%%%%%%%%%%%%%%%%%%%%%%%%
\subsubsection{Stage 3: Segmentation Correction and Refinement}
\label{sub-sub-sec:model_stage_3}

This stage is composed of two individual models. The first model incorporates the segmentation probability map from the models trained in stage 2 as well as the CT volume from each of the phases as inputs. This model is trained on patches of the same size as the patches used in stage 2. The model has three individual encoder branches for each of the three phases followed by feature fusion through concatenation and a convolutional block, which is then followed by a single decoder branch to generate the overall segmentation map. Each of the encoder branches has two input channels. The first channel is the CT volume patch from one of the phases. The second input channel is the mean of the segmentation probability map output of the model trained on the same phase and the model trained on all phases from stage 2. This model structure is shown in Fig. \ref{fig:framework}.  

The second model works on refining the segmentation map by lesion. For each of the lesions predicted by the first model in stage 3, this model uses an encoder-decoder structure as shown in Fig. \ref{fig:framework} to refine the segmentation for that lesion. To identify and separate lesions, we use morphological connectivity to identify each of the lesions, then create a bounding box around this lesion with a margin of 20\% as input to this model. The segmentation map is then updated based on the segmentation output for each of the lesions individually to create the overall final segmentation map.

\begin{figure*}[h]
\vspace{0.05cm}
\hspace{-1mm}
\begin{minipage}[b]{.166\linewidth}
  \centering
  \centerline{\includegraphics[height=3cm, width=1\linewidth]{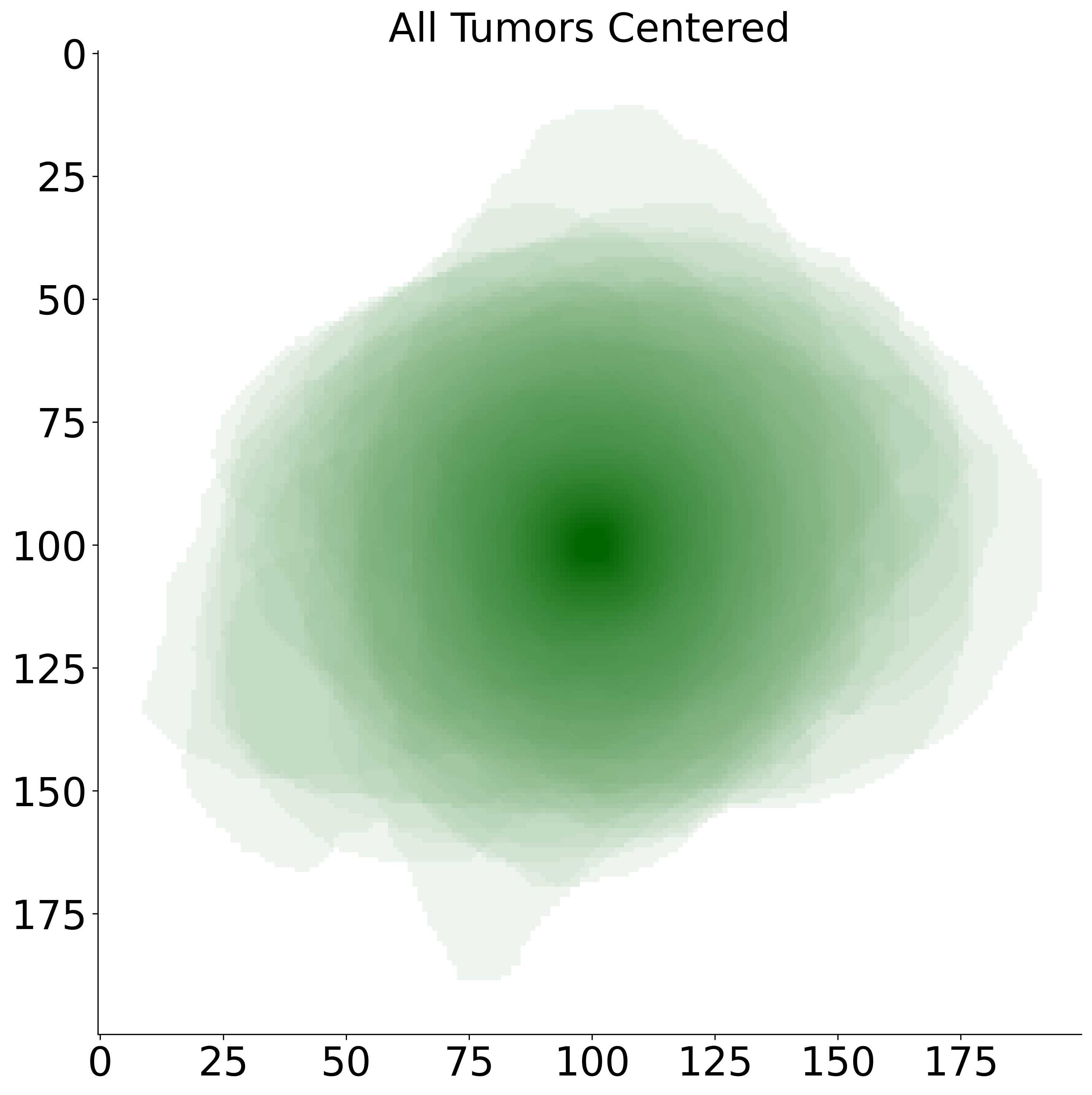}}
  \vspace{-0.05cm}
  \centerline{\footnotesize(a)}
\end{minipage}
\hspace{-1.25mm}
\begin{minipage}[b]{.166\linewidth}
  \centering
  \centerline{\includegraphics[height=3cm, width=1\linewidth]{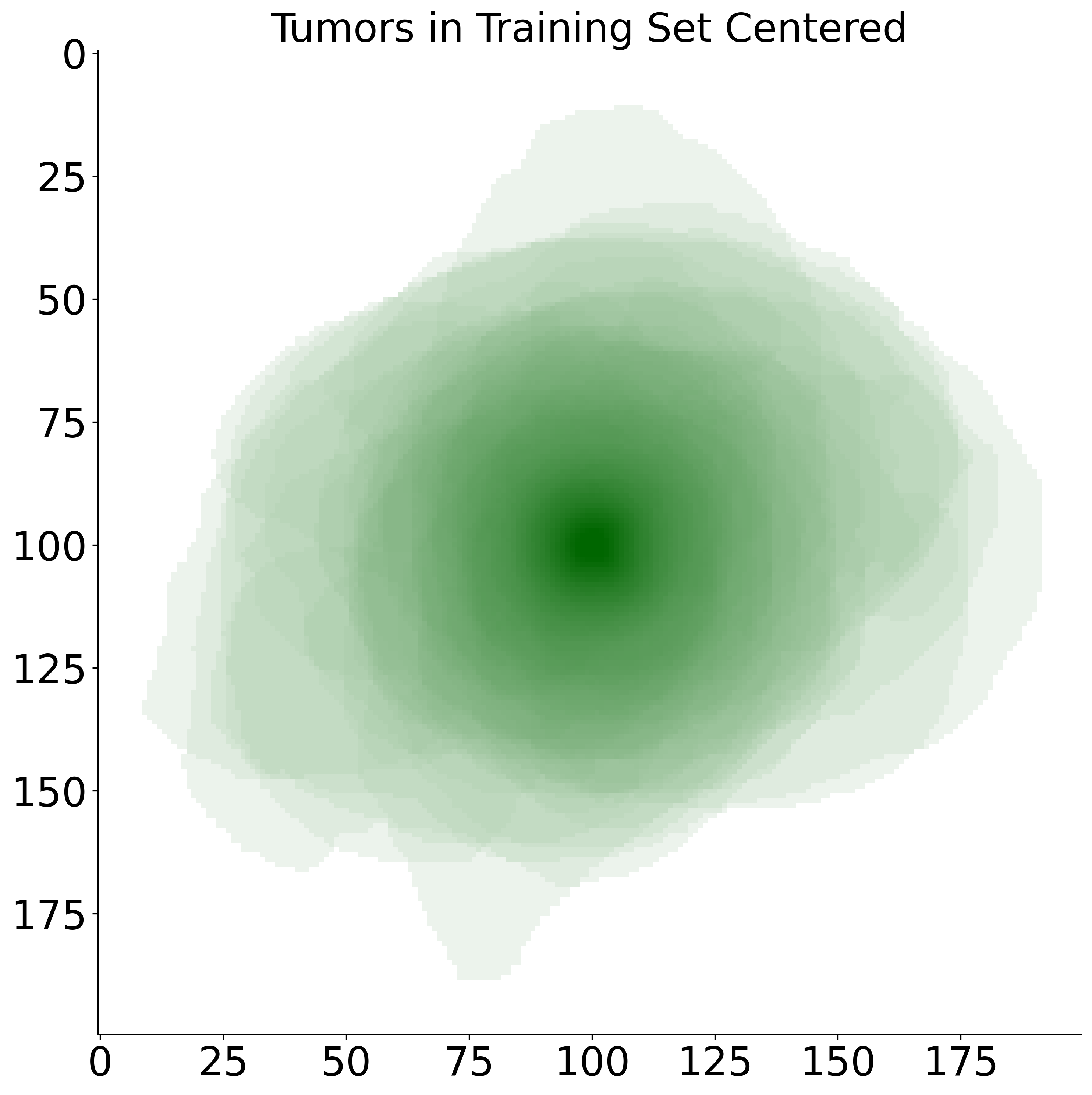}}
  \vspace{-0.05cm}
  \centerline{\footnotesize(b)}
\end{minipage}
\hspace{-1.25mm}
\begin{minipage}[b]{.166\linewidth}
  \centering
  \centerline{\includegraphics[height=3cm, width=1\linewidth]{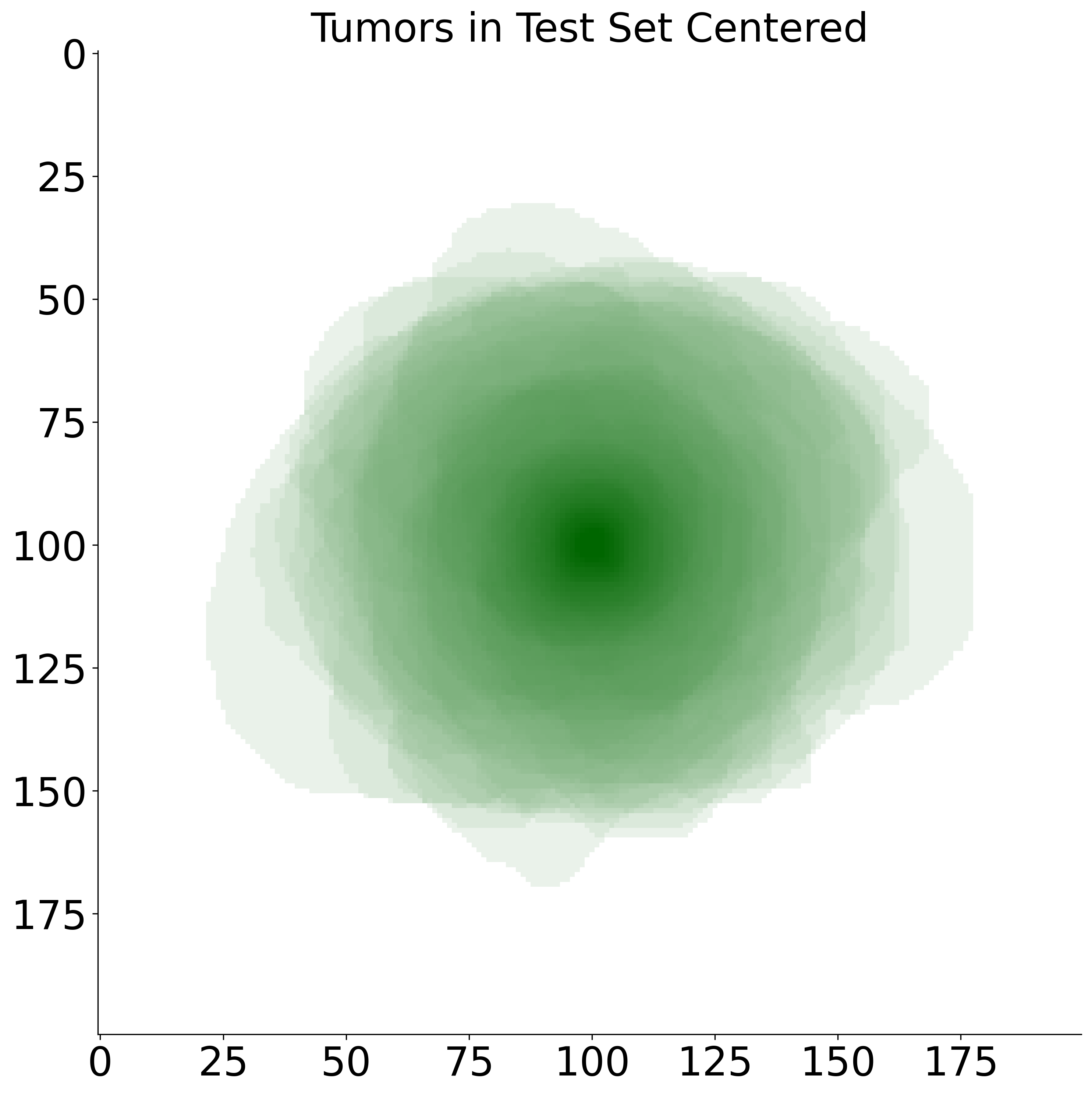}}
  \vspace{-0.05cm}
  \centerline{\footnotesize(c)}
\end{minipage}
\hspace{-1.25mm}
\begin{minipage}[b]{.166\linewidth}
  \centering
  \centerline{\includegraphics[height=3cm, width=1\linewidth]{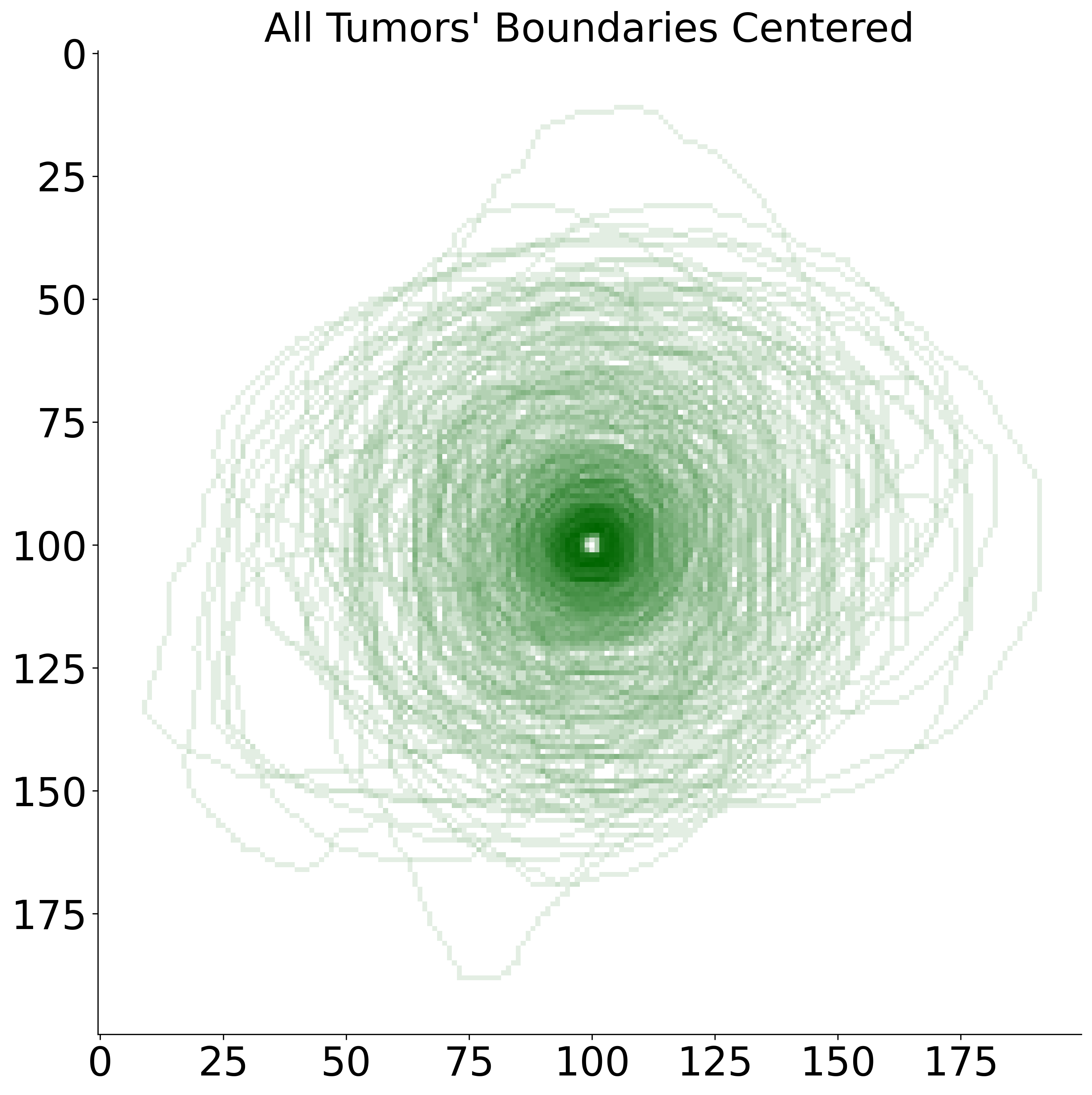}}
  \vspace{-0.05cm}
  \centerline{\footnotesize(d)}
\end{minipage}
\hspace{-1.25mm}
\begin{minipage}[b]{.166\linewidth}
  \centering
  \centerline{\includegraphics[height=3cm, width=1\linewidth]{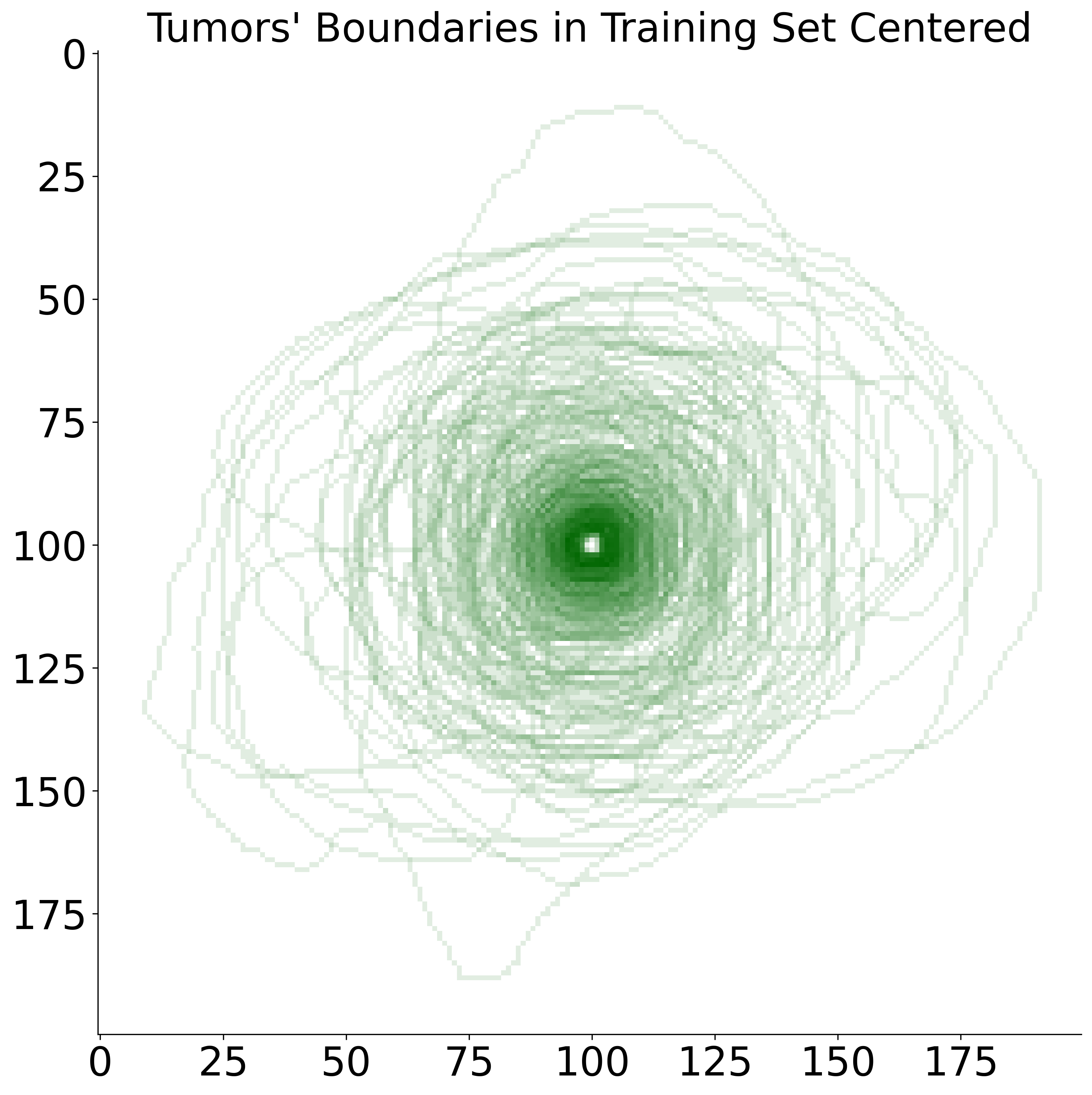}}
  \vspace{-0.05cm}
  \centerline{\footnotesize(e)}
\end{minipage}
\hspace{-1.25mm}
\begin{minipage}[b]{.166\linewidth}
  \centering
  \centerline{\includegraphics[height=3cm, width=1\linewidth]{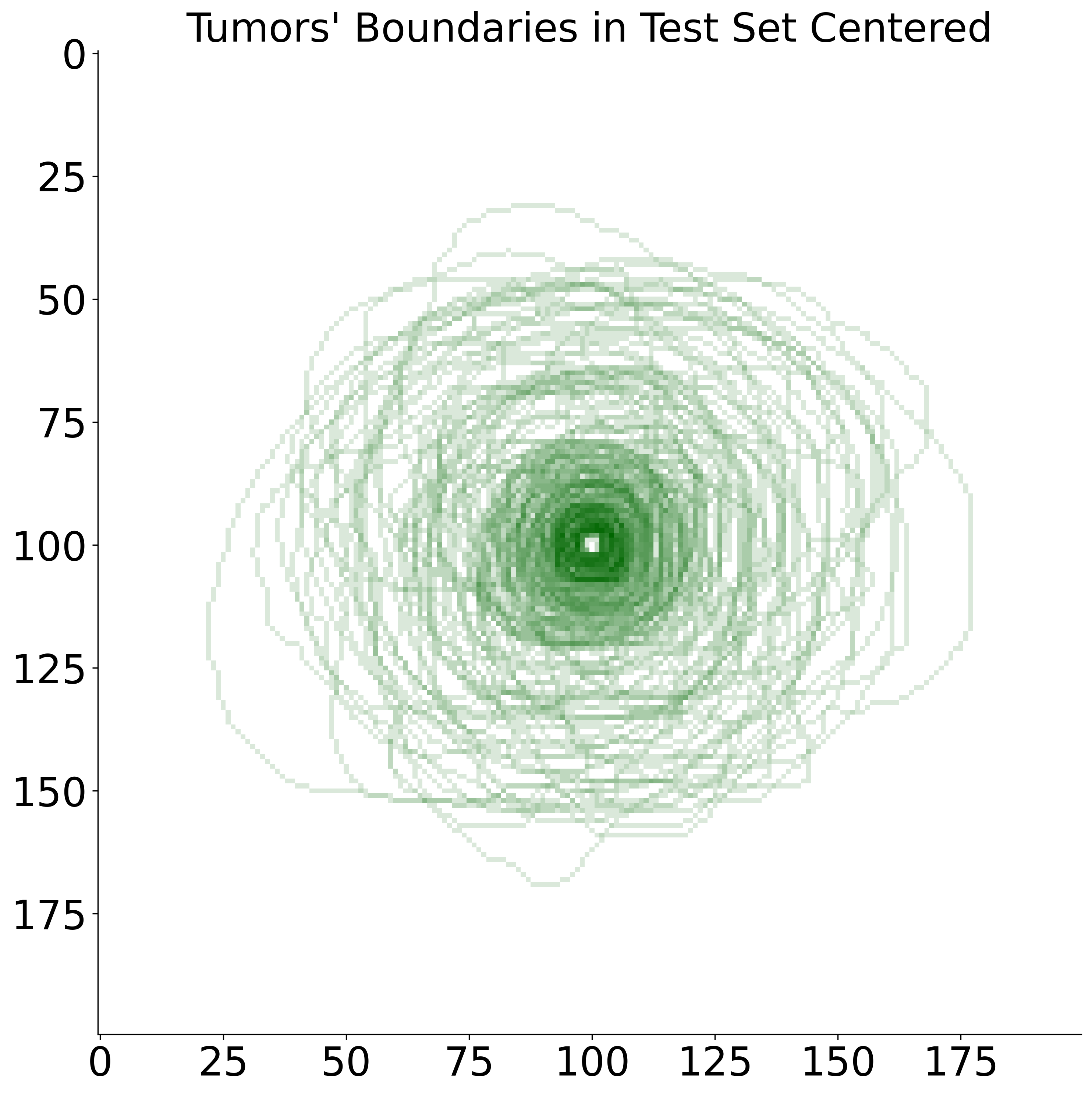}}
  \vspace{-0.05cm}
  \centerline{\footnotesize(f)}
\end{minipage}
\\
\vspace{-0.5cm}
\caption{The largest axial slice of each liver lesion in the whole dataset (a), the training set (b), and the test set (c) overlaid onto one image at a scale of 1 mm. The boundary of these lesions in the same axial slice for the whole dataset (d), training set (e), and test set (f).}
\label{fig:tumor_overlay}
\vspace{-0.2cm}
\end{figure*}

 %%%%%%%%%%%%%%%%%%%%%%%%%%%%%%%%%%%%%%%%%%%%%%%%%%%
\subsection{Model Training and Segmentation Refinement}
\label{sub-sec:model_train}
All models in the different stages of the proposed framework are trained using a compound loss function of two loss functions. The first is the binary cross-entropy (BCE) loss function, which promotes matching the predicted mask map with the ground truth on a voxel by voxel basis globally over the whole mask without explicit consideration for overlap or object-based penalization. For a predicted mask ($X$) and ground truth mask ($Y$), both of of size $H \times W \times D$, the loss is defined for each training example as:
\begin{equation}
\label{eq:bce_mask}
\begin{gathered}
\mathcal{L}_{bce}(X,Y) = - \frac{1}{H \times W \times D} \sum_{i=1}^{H}  \sum_{j=1}^{W} \sum_{k=1}^{D}  \ell\left(x_{ijk} , y_{ijk}\right),
\end{gathered}
\end{equation}
where the predicted mask $x$ and ground truth $y$ at location $(i,j, k)$ are used to calculate the loss $\ell\left(x_{i,j,k} , y_{i,j,k}\right)$ at each of the voxels:
\begin{equation}
\label{eq:bce_element}
\begin{gathered}
\ell\left(x_{ijk} , y_{ijk}\right) =  w_{c} y_{ijk} \log{\sigma(x_{ijk})} + (1-y_{ijk}) \log{\sigma(x_{ijk})}.
\end{gathered}
\end{equation}

\noindent In (\ref{eq:bce_element}), $\sigma(x)$ is the sigmoid function defined as $\sigma(x) = 1 / (1+exp(-x))$, and $i = 1, 2, ..., H$, $j = 1, 2, ..., W$, and $k = 1, 2, ..., D$. $\sigma(x)$ maps the predicted voxels onto a probability space of predictions; its value indicates an object if it is larger than or equal to a threshold, and background otherwise (this threshold is usually set to 0.5, the median between 0 and 1). The second loss function is the Dice loss, which uses the Dice coefficient between the predicted and ground truth mask. The Dice coefficient promotes overlap matching and provides an explicit localized and object-based penalization. The Dice coefficient ($D_{c}$) is defined as \citep{dice_paper_segmentation}:

\begin{equation}
\label{eq:dice_coeff}
\begin{gathered}
D_{c}(\widehat{Y}, Y) =  \frac{2\sum(\widehat{Y} \odot Y)}{\sum_{i,j,k=1}^{H,W,D}\widehat{y}_{ijk} + \sum_{i,j, k=1}^{H,W,K}y_{ijk}}
\end{gathered}
\end{equation}
\noindent $\odot$ is element-wise multiplication and $\widehat{y}_{ijk} = \sigma(x_{ijk})$. The loss based on the dice coefficient ($D_{c}$) promotes higher $D_{c}$ values while countering lower ones. It is defined as:
\begin{equation}
\label{eq:dice_loss}
\begin{gathered}
\mathcal{L}_{D_{c}}(X,Y) = 1 - D_{c}(\widehat{Y}, Y).
\end{gathered}
\end{equation}
\noindent The total compund loss function is defined as the weighted sum of the two losses:
\begin{equation}
\label{eq:tot_loss}
\begin{gathered}
\mathcal{L}_{obj}(X,Y) = \alpha_{b}\mathcal{L}_{bce}(X,Y) + \alpha_{d}\mathcal{L}_{D_{c}}(X,Y),
\end{gathered}
\end{equation}
\noindent where the coefficients $\alpha_{b}$ and $\alpha_{d}$ control the contribution of each loss to the total loss function. In our implementation we chose $\alpha_{b} = \alpha_{d} = 1$. 

\subsubsection{Stage 3 Model Training}
\label{subsub-sec:model_train_stage_3}

In stage 3, the first model processes inputs from the three phases of CT scans and the segmentation probability maps to generate an initial segmentation map. Each encoder branch takes a CT volume patch and the averaged segmentation probability maps as input. For each phase $p \in \{A,D,V\}$, where $A$ stands for arterial, $D$ for delayed, and $V$ for venous. The input to the model is formed as follows:
\begin{equation}
\label{eq:in_stage_3_mod_1}
\begin{gathered}
I_p = \mathfrak{C}_C(V_p, \frac{1}{2}(X_p + X_{3p})),
\end{gathered}
\end{equation}
\noindent where $\mathfrak{C}_c$ represents channel-wise concatenation, $V_p$ the CT volume patch from each of the phases, and $X_p$ and $X_{3p}$ are the segmentation probability maps from the models trained on this phase and three phases in stage 2, respectively. Each of the encoders generates a feature map ($E_p$) from each of the phases individually, which we define as $E_p = \mathfrak{F}_E (I_p)$ where $\mathfrak{F}_E $ represents the encoder feature extraction operations outlined in Fig. \ref{fig:framework}. The overall output segmentation map from this model is then computed as follows:
\begin{equation}
\label{eq:in_stage_3_mod_1_out}
\begin{gathered}
X_R = \mathfrak{F}_D\mathfrak{C}_C(E_A,E_D,E_V).
\end{gathered}
\end{equation}

In (\ref{eq:in_stage_3_mod_1_out}), $\mathfrak{F}_D$ represents the decoder operations, and $\mathfrak{C}_c$ channel-wise concatenation. $E_A$, $E_D$ and $E_V$ are the encoder feature outputs of the arterial, delay, and venous phases. For the second model in stage 3, the input to the model is formed by isolating each of the lesions in $X_R$ using morphological connectivity and identifying the width, height, and depth of the lesion. We then crop a bounding box that is 20\% larger than the lesion span in each of these dimensions from the CT volume $V$. Finally we iteratively update $X_R$ with the refined segmentation map of each of the lesions as follows: 
\begin{equation}
\label{eq:in_stage_3_mod_2_out}
\begin{gathered}
X_{R} = \sum_{l=1}^{L} \mathfrak{F}_{M2}(V_{l}), 
\end{gathered}
\end{equation}

\noindent where $\mathfrak{F}_{M2}$ represents the the second model operations and $V_{l}$ the cropped region from the three phase CT volume for each of the lesions $l = 1, 2, ..., L$.

%%%%%%%%%%%%%%%%%%%%%%%%%%%%%%%%%%%%%%%%%%%%%%%%%%%
%%%%%%%%%%%%%%%%%%%%%%%%%%%%%%%%%%%%%%%%%%%%%%%%%%%

\section{Experiments And Results}
\label{sec:exp}
\subsection{Datasets}
\label{sub-sec:data}

We evaluated our model on 2 different datasets. The first dataset is the main dataset we used to test our approach. It is a clinical dataset that was developed by researchers and clinicians at VinBrain, JSC and the University Medical Center at Ho Chi Minh City. This dataset contains contrast-enhanced 3-phase (arterial, delay, and venous) CT scans of the liver from 354 subjects. The dataset was annotated and rated by two radiologists with extensive experience in liver oncology. Each of the axial scans in the dataset has a resolution of $512 \times 512$ pixels at a physical spacing that ranges from 0.5 mm to 0.84 mm with an average of 0.66mm. Slice thickness in the dataset ranges from 0.5mm to 1mm with an average of 0.9 mm. The average number of lesions in each of the scans is 2.2 while the maximum is 11 and the minimum is 1. Lesions are of significantly varying sizes with lesions as large as 129 mm and as small as 2.7 mm in diameter present within the dataset scans. Masks of these lesions and their boundaries overlaid using their largest axial slice are shown in Fig. \ref{fig:tumor_overlay} to demonstrate the variability in lesion sizes and boundaries within the dataset.  The scans from the arterial phase and delay phase were registered on the venous phase for each of the subjects. This dataset have lesions with varying shapes, sizes, and semantics with respect to healthy liver tissue, which forms a reasonable challenge to test the performance of the proposed framework.

In addition to the main dataset, we test our approach on the BraTS19 \citep{menze2014multimodal,bakas2017advancing,bakas2018identifying} dataset to evaluate its ability to extend its performance improvements to other multi-phase and multi-contrast datasets beyond liver lesion segmentation. The BraTS19 dataset contains MRI scans of the brain from 484 subjects for the purpose of brain tumor segmentation. For each subject, the dataset contains 4 multi-contrast scans, which are: a) native T1, b) post-contrast T1-weighted, c) T2-weighted, and d) T2 Fluid Attenuated Inversion Recovery (FLAIR). The dataset was annotated by 1 to 4 raters and these annotations were approved by experienced neuroradiologists. All the scans are distributed after they have been pre-processed through co-registration to the same anatomical template atlas, interpolated to the same spatial resolution of 1 mm, and skull-stripped.

\begin{table*}[t]
\small
    \caption{The proposed framework liver lesion segmentation performance on the multi-phase CT dataset. Section I outlines the results of just the segmentation model (stage 2 for our framework). For each of the arterial, delay, and venous phases the input number of channels is 1 while for 3-Phase, the number of input channels is 3 (arterial, delay and venous stacked). Section II outlines the performance of the overall proposed framework with our proposed segmentation model in stage 2 as well as the nnUNet and MedNext models. Best and 2\textsuperscript{nd} best results are boldfaced for each category in Section I while only the best is boldfaced in Section II. All metrics are in the range 0 to 100. Values in parenthesis represent the standard deviation across subjects.}
    \vspace{2mm}
    \centering
    \begin{tabular*}{\textwidth}{@{\extracolsep{\fill}}p{1.5cm}p{2.2cm}cccccc}
    \toprule
    & & Global & \multicolumn{4}{c}{By Subject} & Surface  \\ 
    \cmidrule(l{1mm}r{1mm}){3-3} \cmidrule(l{1mm}r{1mm}){4-7} \cmidrule(l{1mm}r{1mm}){8-8}
    Phase & Model & Dice & Dice & IoU & Recall & Precision & Dice \\
    \midrule
    \multicolumn{8}{c}{Section I: Segmentation Model Results} \\
    \midrule
    \multirow{5}{*}{Arterial}
        & SwinUNetR 
            & 71.5
            & 60.3 ($\pm$ 28.5) & 48.6 ($\pm$ 26.8)
            & 62.4 ($\pm$ 30.3) & 68.6 ($\pm$ 30.3)
            & 43.8 \\
        & Model Gnesis 
            & 75.8
            & 63.1 ($\pm$ 26.1) & 50.8 ($\pm$ 25.0)
            & 72.4 ($\pm$ 26.0) & 63.2 ($\pm$ 28.0)
            & 45.5 \\
        & nnUNet 
            & 77.2
            & 65.7 ($\pm$ 26.5) & 54.0 ($\pm$ 25.9)
            & 65.1 ($\pm$ 29.0) & \textbf{76.7} ($\pm$ 25.8)
            & 51.1 \\
        & MedNext 
            & \textbf{80.7}
            & \textbf{69.5} ($\pm$ 23.7) & \textbf{57.6} ($\pm$ 24.2)
            & \textbf{75.9} ($\pm$ 23.0) & 71.6 ($\pm$ 26.7)
            & \textbf{54.0} \\
        & Ours 
            & \textbf{80.8}
            & \textbf{69.8} ($\pm$ 24.0) & \textbf{58.0} ($\pm$ 24.2)
            & \textbf{75.7} ($\pm$ 23.4) & \textbf{72.1} ($\pm$ 26.8)
            & \textbf{54.3} \\
    \midrule
    \multirow{5}{*}{Delay}
        & SwinUNetR 
            & 79.5
            & 61.6 ($\pm$ 28.6) & 50.0 ($\pm$ 26.8)
            & 66.5 ($\pm$ 30.8) & 66.6 ($\pm$ 30.4)
            & 44.1\\
        & Model Genesis 
            & 78.5
            & 62.4 ($\pm$ 28.2) & 50.8 ($\pm$ 26.7)
            & \textbf{67.6} ($\pm$ 29.4) & 66.4 ($\pm$ 30.2)
            & 45.7 \\
        & nnUNet 
            & 81.8
            & 64.3 ($\pm$ 29.5) & 53.3 ($\pm$ 27.6)
            & 64.4 ($\pm$ 31.4) & 73.2 ($\pm$ 29.1)
            & 48.8 \\
        & MedNext 
            & \textbf{82.7}
            & \textbf{67.3} ($\pm$ 27.1) & \textbf{56.0} ($\pm$ 26.1)
            & 66.7 ($\pm$ 28.6) & \textbf{76.8} ($\pm$ 25.6)
            & \textbf{52.6} \\
        & Ours 
            & \textbf{83.0}
            & \textbf{67.4} ($\pm$ 27.0) & \textbf{56.1} ($\pm$ 26.1)
            & \textbf{66.9} ($\pm$ 28.8) & \textbf{76.8} ($\pm$ 25.5)
            & \textbf{52.8} \\
    \midrule
    \multirow{5}{*}{Venous}
        & SwinUNetR 
            & 80.0
            & 63.6 ($\pm$ 26.6) & 51.6 ($\pm$ 25.7)
            & 66.3 ($\pm$ 28.3) & 69.9 ($\pm$ 29.0)
            & 47.5 \\
        & Model Genesis 
            & 79.4
            & 63.3 ($\pm$ 29.7) & 52.3 ($\pm$ 27.8)
            & 68.0 ($\pm$ 30.8) & 67.4 ($\pm$ 30.7)
            & 47.8 \\
        & nnUNet 
            & \textbf{82.9}
            & 65.5 ($\pm$ 29.2) & 54.6 ($\pm$ 27.8)
            & 65.0 ($\pm$ 30.8) & \textbf{73.6} ($\pm$ 29.6)
            & 51.5 \\
        & MedNext 
            & 82.1
            & \textbf{67.8} ($\pm$ 26.5) & \textbf{56.3} ($\pm$ 26.2)
            & \textbf{69.9} ($\pm$ 27.6) & 73.4 ($\pm$ 28.2)
            & \textbf{54.0} \\
        & Ours 
            & \textbf{82.2}
            & \textbf{67.8} ($\pm$ 26.4) & \textbf{56.5} ($\pm$ 26.1)
            & \textbf{69.8} ($\pm$ 27.4) & \textbf{73.5} ($\pm$ 28.0)
            & \textbf{54.1} \\
    \midrule
    \multirow{6}{*}{3-Phase}
        & SwinUNetR 
            & 81.6
            & 68.2 ($\pm$ 23.2) & 55.8 ($\pm$ 23.5)
            & 68.1 ($\pm$ 25.0) & 78.0 ($\pm$ 23.6)
            & 53.1 \\
        & Model Genesis 
            & 80.8
            & 70.7 ($\pm$ 22.4) & 58.7 ($\pm$ 22.8)
            & 73.4 ($\pm$ 23.5) & 76.6 ($\pm$ 23.2)
            & 57.6 \\
        & OctopusNet 
            & 75.5
            & 67.2 ($\pm$ 22.7) & 54.4 ($\pm$ 22.5)
            & 64.2 ($\pm$ 24.3) & \textbf{81.2} ($\pm$ 23.5)
            & 52.4 \\
        & nnUNet 
            & 83.7
            & 73.4 ($\pm$ 22.4) & 62.0 ($\pm$ 23.0)
            & 74.8 ($\pm$ 24.2) & 79.4 ($\pm$ 22.0)
            & 61.8 \\
        & MedNext 
            & \textbf{83.9}
            & \textbf{75.1} ($\pm$ 20.1) & \textbf{63.6} ($\pm$ 21.1)
            & \textbf{76.8} ($\pm$ 22.6) & 80.3 ($\pm$ 19.1)
            & \textbf{64.1} \\
        & Ours 
            & \textbf{84.1}
            & \textbf{75.5} ($\pm$ 19.8) & \textbf{63.9} ($\pm$ 20.8)
            & \textbf{76.8} ($\pm$ 22.1) & \textbf{80.9} ($\pm$ 18.9)
            & \textbf{64.4} \\
    \midrule
    \multicolumn{8}{c}{Section II: Overall Framework Results} \\
    \midrule
    \multirow{3}{*}{Overall}
        & Ours (nnUNet) 
            & 84.2
            & 74.1 ($\pm$ 22.1) & 62.8 ($\pm$ 22.7)
            & 77.7 ($\pm$ 22.8) & 77.3 ($\pm$ 22.2)
            & 62.7 \\
        & Ours (MedNext)  
            & 84.6
            & 75.8 ($\pm$ 18.9) & 63.8 ($\pm$ 20.9)
            & 79.1 ($\pm$ 19.5) & 78.6 ($\pm$ 20.8)
            & 64.5 \\
        & Ours (Ours)
            & \textbf{85.1}
            & \textbf{76.3} ($\pm$ 18.5) & \textbf{64.6} ($\pm$ 20.3)
            & \textbf{79.1} ($\pm$ 20.0) & \textbf{80.0} ($\pm$ 19.4)
            & \textbf{65.0} \\
    % Repeat the above line for each entry
    \bottomrule
    \end{tabular*}
    \label{tab:results_main}
    \vspace{-0.3cm}
\end{table*}

\subsection{Experimental Setup and Data Preperation}
\label{sub-sec:setup}

For each of the datasets, we trained the models in the proposed framework from randomly initiated weights. For the liver lesion dataset the scans were split into 200 for training and 154 for testing while for the BraTS dataset the split was 387 for training and 97 for testing. For the liver lesion dataset, the liver was segmented using a model trained on the LiTS dataset and scans with only the liver region were used for the training and testing of the proposed lesion segmentation approach. The BraTS dataset is already pre-processed with the organ of interest (brain) isolated. We used a spatial resolution of 1 mm\textsuperscript{3} for both datasets and patches of size $128 \times 128 \times 128$ voxels for the models in stage 2 and 3, and patches of size $64 \times 64 \times 64$ voxels for the model in stage 1. The CT scans intensity values which are represented by the Hounsfield units were clipped to the range [-200, 200] before normalization. We compared the proposed approach performance to four 3D segmentation networks, which are the current leading models across different medical segmentation tasks. These models are the SwinUnetR \citep{hatamizadeh2021swin}, Model Genesis \citep{zhou2021models}, nnUNet \citep{isensee2021nnu}, and MedNext \citep{roy2023mednext}. We also compared our approach to OctopusNet, which is specifically designed for multi-phase and multi-contrast medical images. Furthermore, we incorporate the two best performing models out of these four, which are the nnUNet and MedNext models, as the stage 2 models in our framework to demonstrate the ability of the overall framework we propose to improve the segmentation output of these models. The models were trained for 800 to 1200 epochs depending on the learning rate that is suitable for the model, which ranged from $1e^{-4}$ to $1e^{-2}$. All the models were trained using the AdamW \citep{loshchilov2017decoupled} optimizer and loss function defined in (\ref{eq:tot_loss}), except for the nnUNet and Model Genesis models, which are trained using the stochastic gradient descent optimizer as it is the recommended optimizer for both models. At each iteration within an epoch, we select two patches randomly from the image volume using a weighted sampling scheme that gives a 50\% higher weight to the probability of sampling a patch containing a lesion. The second model in stage 3 was trained using a single patch at a time as the patches were of varying size depending on the lesion size.

\begin{figure*}[hbt!]
\vspace{0.05cm}
\begin{minipage}[b]{.99\linewidth}
  \centering
  \centerline{\includegraphics[height=3.6cm, width=0.99\linewidth]{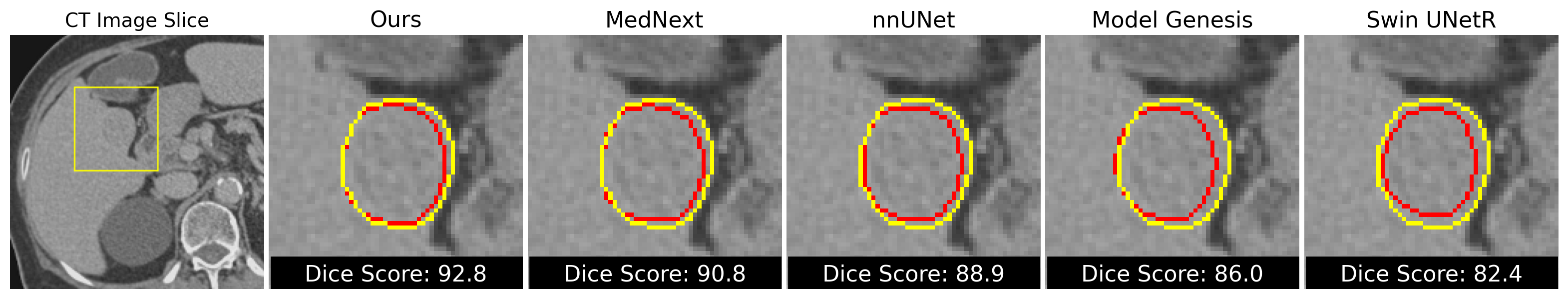}}
\end{minipage}
\\
\begin{minipage}[b]{.99\linewidth}
\vspace{-1.5mm}
  \centering
  \centerline{\includegraphics[height=3.15cm, width=0.99\linewidth, trim = {0cm 0cm 0cm 0.8cm}, clip]{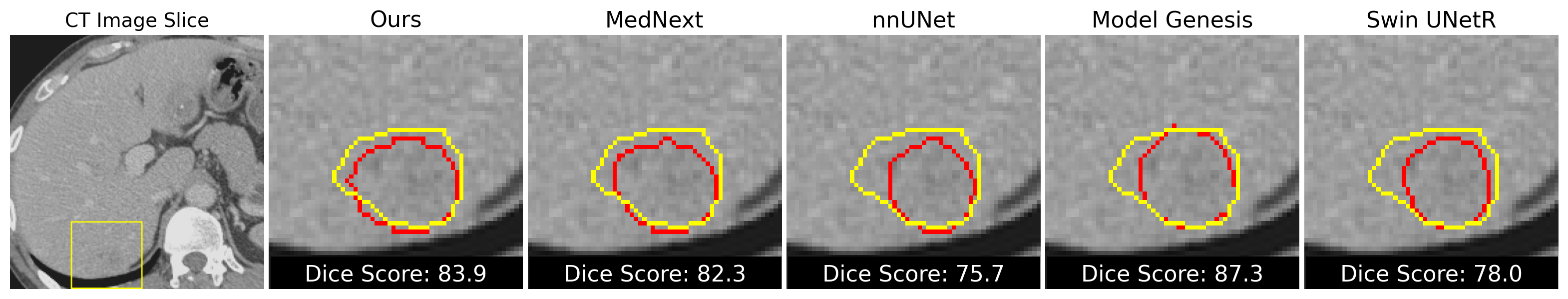}}
\end{minipage}
\\
\begin{minipage}[b]{.99\linewidth}
\vspace{-1.5mm}
  \centering
  \centerline{\includegraphics[height=3.15cm, width=0.99\linewidth, trim = {0cm 0cm 0cm 0.8cm}, clip]{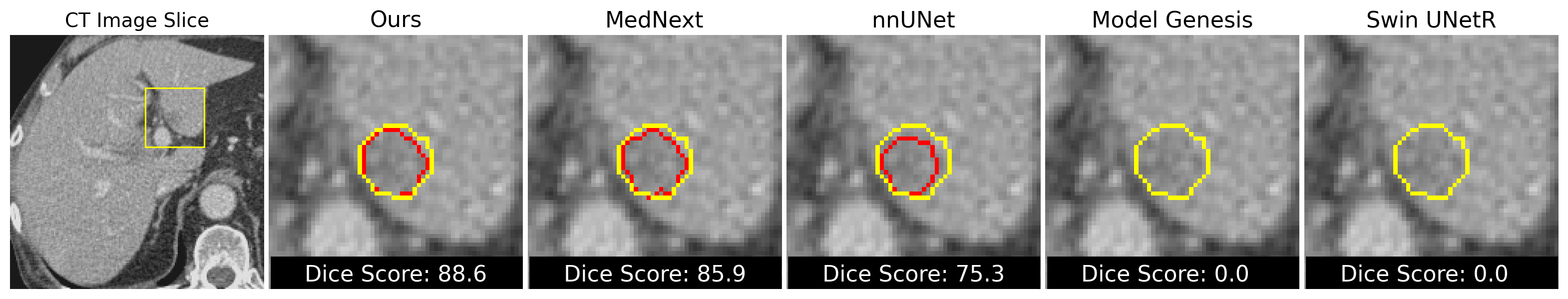}}
\end{minipage}
\\
\begin{minipage}[b]{.99\linewidth}
\vspace{-1.5mm}
  \centering
  \centerline{\includegraphics[height=3.15cm, width=0.99\linewidth, trim = {0cm 0cm 0cm 0.8cm}, clip]{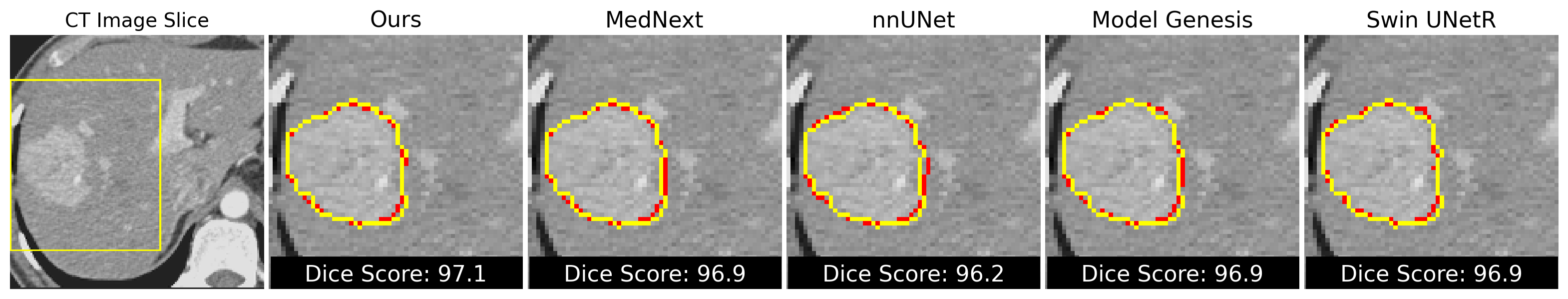}}
\end{minipage}
\\
\begin{minipage}[b]{.99\linewidth}
\vspace{-1.5mm}
  \centering
  \centerline{\includegraphics[height=3.15cm, width=0.99\linewidth, trim = {0cm 0cm 0cm 0.8cm}, clip]{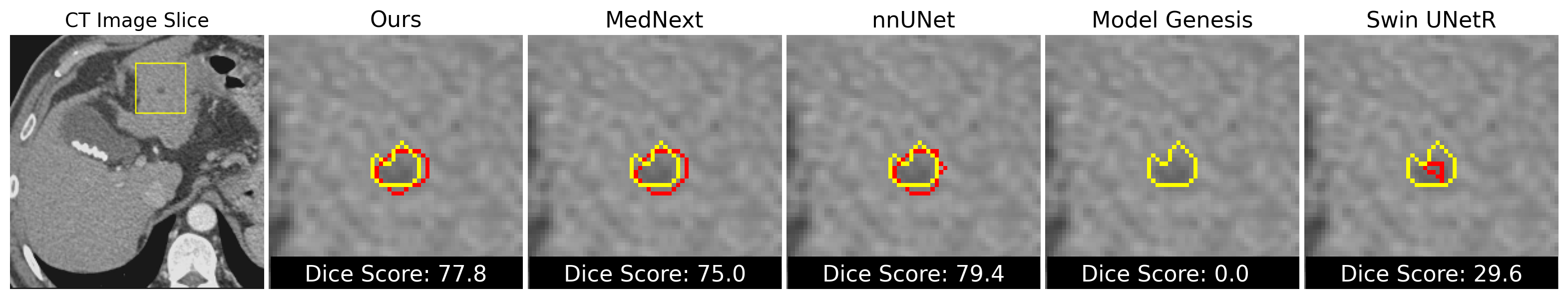}}
\end{minipage}
\\
\vspace{-0.5cm}
\caption{Qualitative comparison of the proposed approach to the other four baseline models. The figure presents a side-by-side assessment of the different approaches' segmentation performance on lesions of different sizes, shapes and intensity characteristics. The first column shows the original CT image slices cropped to the liver region with the region of interest highlighted in a yellow bounding box. The subsequent columns show the cropped bounding box region with the outline of segmentation results from the different approaches (in red) and the ground truth segmentation outline (in yellow). The Dice score for each of the predictions outlined is listed below each image.}
\label{fig:results_qualitative}
\vspace{-0.2cm}
\end{figure*}

\begin{figure*}[h]
\vspace{0.05cm}
\begin{minipage}[b]{.99\linewidth}
  \centering
  \centerline{\includegraphics[height=3.6cm, width=0.99\linewidth]{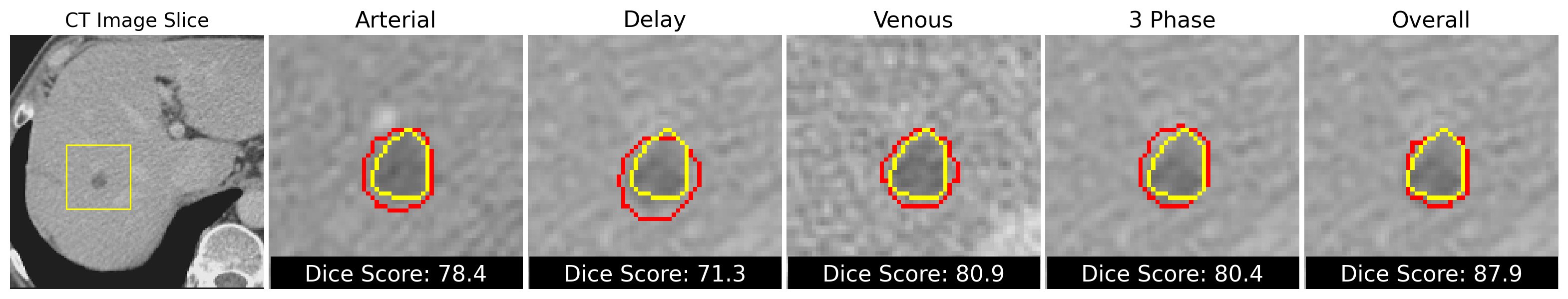}}
\end{minipage}
\\
\begin{minipage}[b]{.99\linewidth}
\vspace{-1.5mm}
  \centering
  \centerline{\includegraphics[height=3.15cm, width=0.99\linewidth, trim = {0cm 0cm 0cm 0.8cm}, clip]{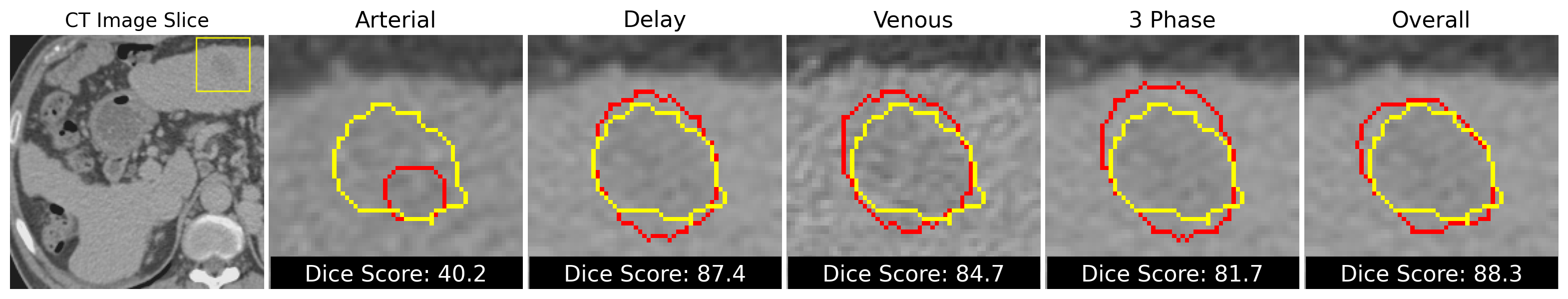}}
\end{minipage}
\\
\begin{minipage}[b]{.99\linewidth}
\vspace{-1.5mm}
  \centering
  \centerline{\includegraphics[height=3.15cm, width=0.99\linewidth, trim = {0cm 0cm 0cm 0.8cm}, clip]{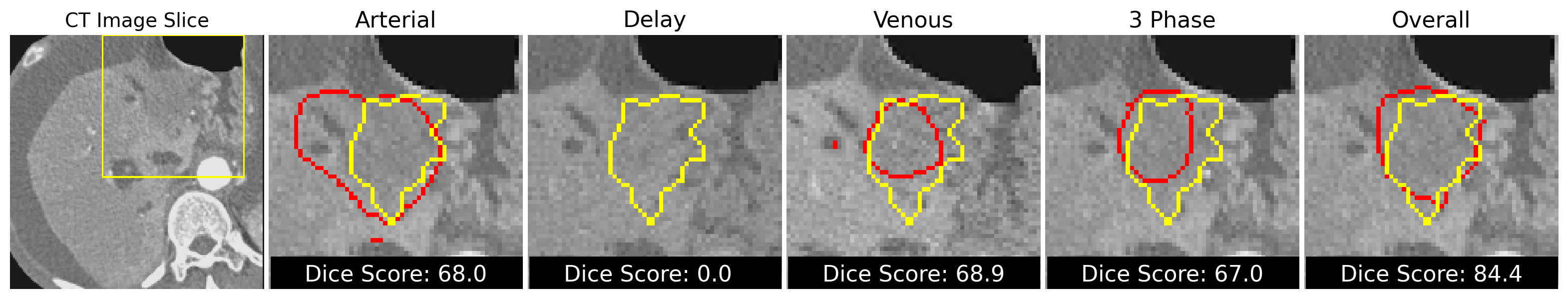}}
\end{minipage}
\\
\begin{minipage}[b]{.99\linewidth}
\vspace{-1.5mm}
  \centering
  \centerline{\includegraphics[height=3.15cm, width=0.99\linewidth, trim = {0cm 0cm 0cm 0.8cm}, clip]{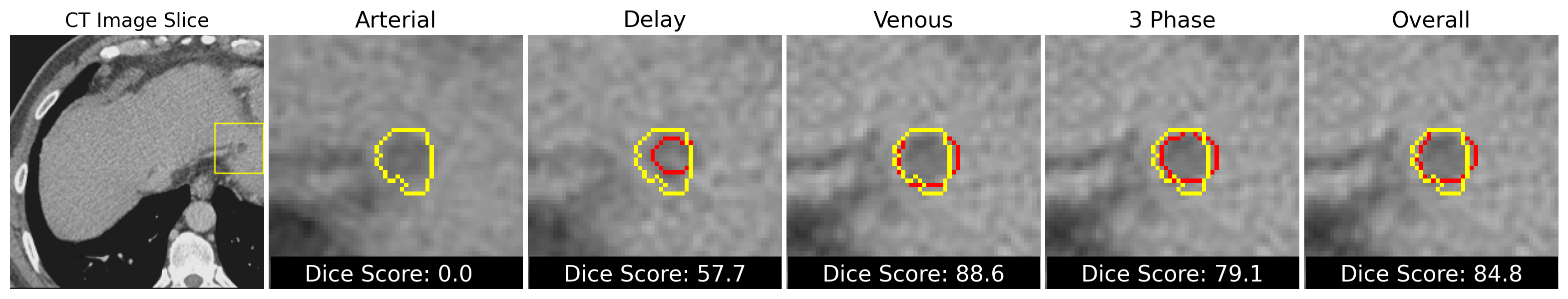}}
\end{minipage}
\\
\vspace{-0.5cm}
\caption{Qualitative demonstration of the proposed framework ability to improve the segmentation outcome versus 3-Phase segmentation models by incorporating leanings from each of the phases individually. The figure presents a side-by-side comparison of the overall framework to the different models trained on each of the phases individually (arterial, delay, and venous) as well as the 3-Phase model. The first column shows the original CT image slices cropped to the liver region with the region of interest highlighted in a yellow bounding box. The subsequent columns show the cropped bounding box region with the outline of segmentation predictions (in red) and the ground truth segmentation outline (in yellow). For each of the models trained on individual phases, the ground truth and predictions are shown on a cropped box region from the corresponding phase slice. The Dice score for each of the predictions outlined is listed below each image.}
\label{fig:results_qualitative_ours}
\vspace{-0.2cm}
\end{figure*}

\subsection{Evaluation and Results}
\label{sub-sec:eval}

\subsubsection{Evaluation Metrics}
\label{subsub-sec:eval_met}

To evaluate the performance of our proposed architecture, we use multiple segmentation, detection and localization metrics. For segmentation, we use the Dice score and the intersection over union (IoU) metrics. Both metrics are considered the benchmark metric used to evaluate detection, and segmentation methods \citep{lin2014microsoft}. The Dice score is defined in (\ref{eq:dice_coeff}) while the IoU metric is calculated as follows:
\begin{equation}
\label{eq:iou}
\begin{gathered}
IoU(\widehat{Y}, Y) =  \frac{\sum(\widehat{Y} \odot Y)}{\sum(\widehat{Y} \, | \, Y)},
\end{gathered}
\end{equation}
\noindent where $\odot$ is element-wise multiplication, $|$ is the element-wise $or$ logical operator, $\widehat{Y}$ is the predicted mask map, and $Y$ is the ground truth. For the Dice score, we measure it globally by aggregating true positives (TP), false positives (FP), and false negatives (FN) over all subjects and then computing the Dice score as follows:
\begin{equation}
\label{eq:dice_tp_fp_fn}
\begin{gathered}
D_c =  \frac{2TP}{2TP+FP+FN}
\end{gathered}
\end{equation}

We also measure the Dice score by subject to capture the variability in model performance across subjects. Furthermore, we evaluate the segmentation recall ($TP / (TP + FN)$) and precision ($TP / (TP + FP)$) by subject as well as the surface Dice score at a tolerance of 1.5mm.

\subsubsection{Overall Segmentation Performance}
\label{subsub-sec:seg_res}
The performance of the proposed approach using the segmentation evaluation metrics outlined in Section \ref{subsub-sec:eval_met} is summarized in Table \ref{tab:results_main}, and is compared to the other 5 benchmark models. In section I of Table \ref{tab:results_main} we compare the performance of segmentation models only (stage 2 for our framework) without the use of the whole framework. In these models, the input is either a single channel 3D image patch for individual phases or a 3-channel 3D image patch for the 3-phase case. In section II of the table we summarize the performance of the whole framework. The proposed approach constantly outperformed the 5 benchmark models, and was able to improve the feature extraction and segmentation of lesions from the different phases leading to a 1.6\% relative Dice score improvement (76.3\% versus 75.1\%) when compared to the best performing benchmark model. This performance improvement translates to better segmentation maps of lesions with better detection and boundaries as shown in Fig. \ref{fig:results_qualitative} and Fig. \ref{fig:results_qualitative_ours}. In Fig. \ref{fig:results_qualitative}, lesions with different characteristics from multiple subjects are shown together with the ground truth and predicted segmentation boundaries from our model and other baseline models. In Fig. \ref{fig:results_qualitative}, we demonstrate the ability of the proposed framework to improve segmentation accuracy when compared to the 3-Phase segmentation models. The first stage of the framework improves the segmentation model Dice score performance across subjects by 0.1\% while stage 3 improves the overall performance  by 0.7\%. The overall framework can also integrate with other segmentation models that can be used instead of our proposed model in stage 2 of the framework. We conducted two experiments with the best two benchmark models. A relative Dice score performance improvements of 1\% were observed for both the nnUNet and MedNext models as shown in Section II of Table \ref{tab:results_main}.  

\begin{figure*}[ht]
\vspace{-1mm}
    \centering
    \includegraphics[width = 0.99\linewidth, height=5.4cm, trim = {0.25cm 0.25cm 0.25cm 0.25cm}, clip]{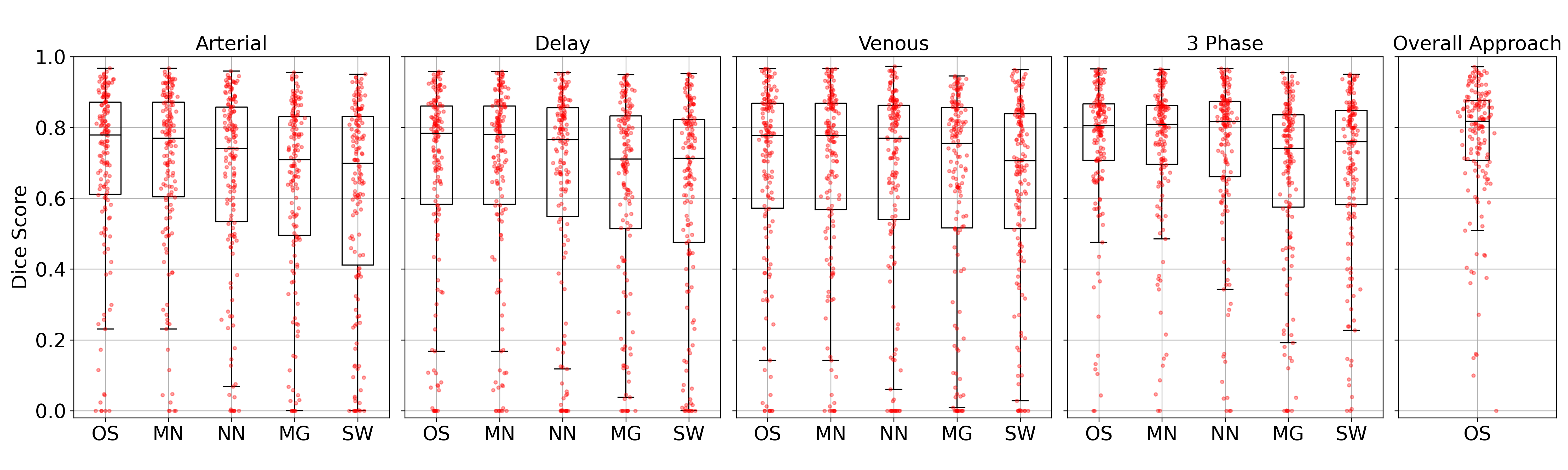}

\begin{minipage}{0.22\linewidth}
  \centering
  \centerline{\hspace{8mm}\footnotesize(a)}\medskip
\end{minipage}
\begin{minipage}{0.21\linewidth}
  \centering
  \centerline{\hspace{5mm}\footnotesize(b)}\medskip
\end{minipage}
\begin{minipage}{0.21\linewidth}
  \centering
  \centerline{\hspace{4mm}\footnotesize(c)}\medskip
\end{minipage}
\begin{minipage}{0.21\linewidth}
  \centering
  \centerline{\hspace{2mm}\footnotesize(d)}\medskip
\end{minipage}
\begin{minipage}{0.1\linewidth}
  \centering
  \centerline{\hspace{2.5mm}\footnotesize(e)}\medskip
\end{minipage}

\vspace{-3mm}
\caption{Boxplots of the proposed segmentation model performance by subject (in terms of Dice score) compared to the other four baseline models when trained on the arterial (a), delay (b) and venous (c) phases individually, and when using the three phases as 3-channel inputs (d). The performance of the overall framework is in (e). For each model as well as the overall framework, the Dice score of each sample from the test set is overlaid on top of the boxplot as a red circle at the vertical location of the corresponding Dice score. Model names keys: OS = Ours, MN = MedNext, NN = nnUNet, MG = Model Genesis, and SW = SwinUNetR.}
\vspace{-4mm}
\label{fig:box_subject}
\end{figure*}

\begin{table}[h]
\footnotesize
    \caption{The proposed framework brain tumor segmentation performance on the BraTS dataset. Similar to Table \ref{tab:results_main}, Section I outlines the results of just the segmentation model while Section II outlines the performance of the overall proposed framework. All metrics are in the range 0 to 100. Values in parenthesis represent the standard deviation across subjects. Best results are boldfaced}
    \vspace{2mm}
    \centering
    \begin{tabular*}{0.99\linewidth}{@{\extracolsep{\fill}}p{0.7cm}p{0.9cm}cccc}
    \toprule
    Phase & Model & Dice & IoU & Recall & Precision \\
    \midrule
    \multicolumn{6}{c}{Section I: Segmentation Model Results} \\
    \midrule
    \multirow{2}{*}{T1}
        & MedNext 
            & 86.6 (8.0) & 77.1 (11.2)
            & 86.4 (10.4) & 88.0 (9.5) \\
        & Ours 
            & 86.7 (7.9) & 77.1 (11.2)
            & 86.6 (10.4) & 87.9 (9.4) \\
    \midrule
    \multirow{2}{*}{T2}
        & MedNext 
            & 88.7 (7.8) & 80.4 (10.9)
            & 87.8 (9.7) & 90.6 (9.2) \\
        & Ours 
            & 88.7 (7.5) & 80.5 (10.8)
            & 88.1 (9.5) & 90.4 (9.3) \\
    \midrule
    \multirow{2}{*}{T1Gd}
        & MedNext 
            & 86.5 (7.8) & 77.0 (11.3)
            & 85.7 (10.3) & 88.6 (9.4) \\
        & Ours 
            & 86.6 (7.9) & 77.1 (11.5)
            & 85.7 (10.5) & 88.7 (9.2) \\
    \midrule
    \multirow{2}{*}{FLAIR}
        & MedNext 
            & 89.6 (6.9) & 81.9 (10.0)
            & 87.8 (10.5) & 92.7 (6.7) \\
        & Ours 
            & 89.8 (6.6) & 82.1 (10.0)
            & 88.4 (10.0) & 92.3 (6.8) \\
    \midrule
    \multirow{2}{*}{4-Phase}
        & MedNext 
            & 90.8 (5.8) & 83.7 (9.1)
            & 89.6 (8.7) & \textbf{92.8} (6.8) \\
        & Ours 
            & 90.9 (5.8) & 83.8 (9.0)
            & 90.0 (8.4) & 92.7 (6.9) \\
    \midrule
    \multicolumn{6}{c}{Section II: Overall Framework Results} \\
    \midrule
    {Overall}
        & Ours 
            & \textbf{91.1} (5.5) & \textbf{84.0} (8.7)
            & \textbf{90.3} (7.9) & 92.7 (6.8) \\
    % Repeat the above line for each entry
    \bottomrule
    \end{tabular*}
    \label{tab:results_brats}
    \vspace{-0.3cm}
\end{table}

\subsubsection{Extension to The BraTS Dataset}
\label{subsub-sec:brats_ext}
On the BraTS dataset, we evaluated the ability of the proposed approach to improve the segmentation of brain tumors when tested on a different imaging modality and a different anatomical structure of interest. We can observe from Table \ref{tab:results_brats} that the proposed approach outperforms the current state-of-the-art model and further reduces segmentation variability across subjects. The performance improvement on the BraTS dataset is not as significant when compared to the liver lesion dataset due to the proximity in performance between models trained on each of the individual contrast images and the model trained using all four contrast images together as a 4-channel input. Nevertheless, The proposed framework reduced the relative Dice score variability across subjects, which is represented by the standard deviation, by 5.2\% when compared to the benchmark model.

\subsubsection{Performance Variability Across Subjects}
\label{subsub-sec:perf_var}
Performance consistency when it comes to lesion segmentation and detection is crucial to maintain similar levels of patient care and reduce bias. We designed our framework to improve the recovery and segmentation of lesions by incorporating learning from individual phase models. This also reduces performance variability across subjects as shown in Table \ref{tab:results_main}. The proposed framework reduced the relative Dice score variability across subjects, which is represented by the standard deviation, by 8\% when compared to the best performing benchmark model (18.5\% versus 20.1\%). 

\begin{figure*}[htb!]
\vspace{-1mm}
\begin{minipage}[b]{.99\linewidth}
  \centering
  \centerline{\includegraphics[height=4.5cm, width=0.99\linewidth, trim = {0cm 0.9cm 0cm 0.5cm}, clip]{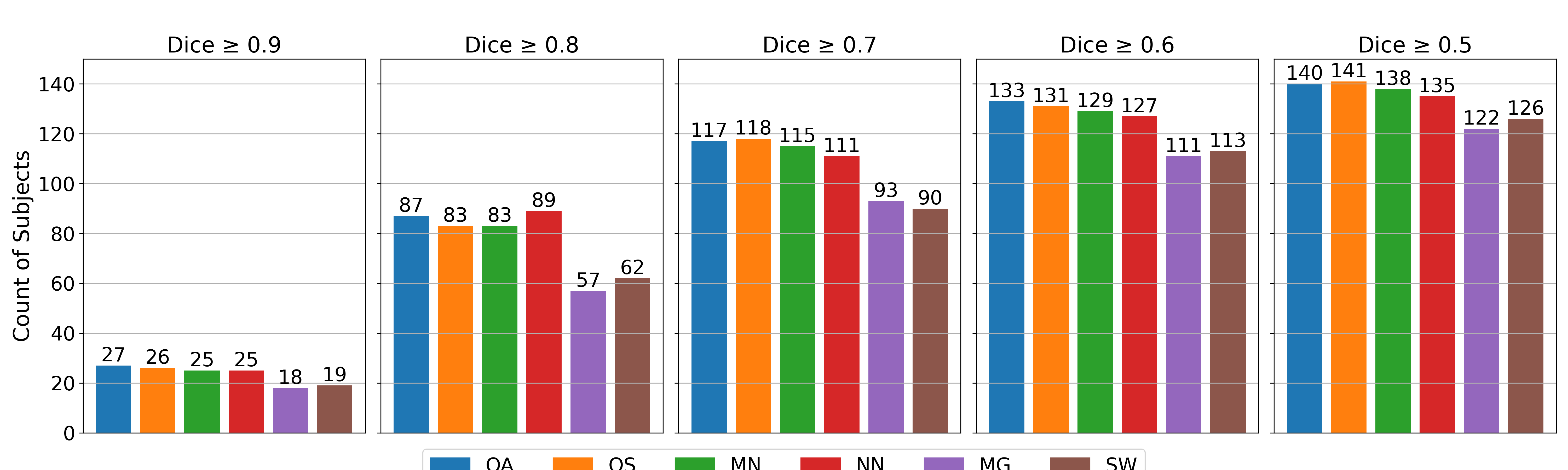}}
\end{minipage}
\\
\centering
\begin{minipage}[b]{.99\linewidth}
  \centering
  \centerline{\includegraphics[height=4.5cm, width=0.99\linewidth, trim = {0cm 0.9cm 0cm 0.5cm}, clip]{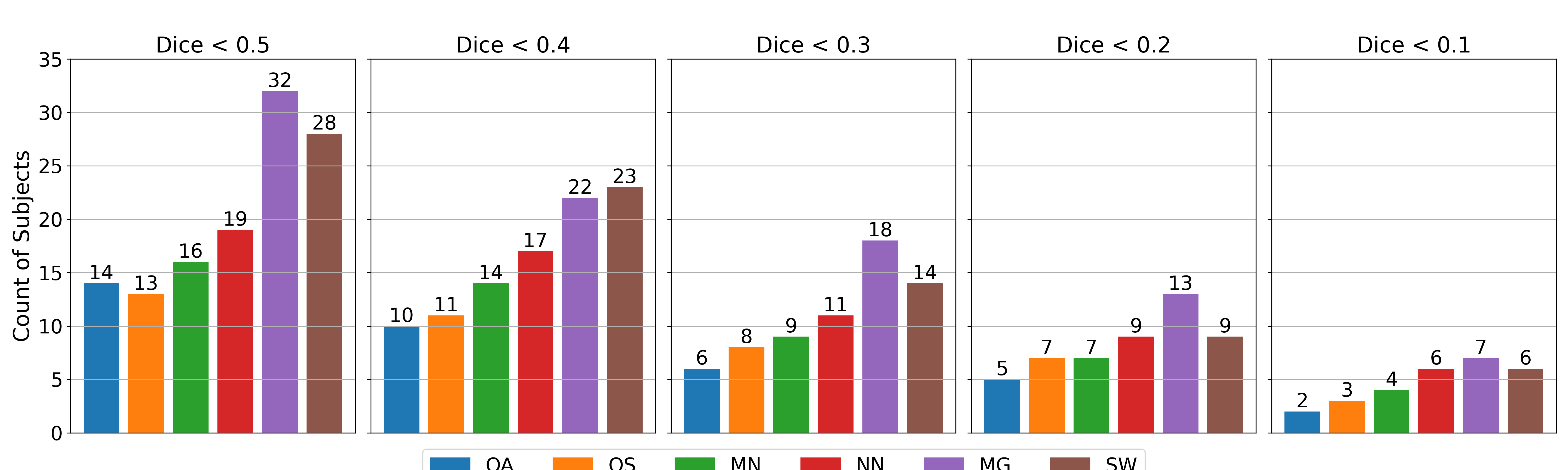}}
\end{minipage}
\begin{minipage}[b]{.99\linewidth}
  \centering
   \centerline{\includegraphics[height=0.6cm, width=0.4\linewidth, trim = {0cm 0cm 0cm 0cm}, clip]{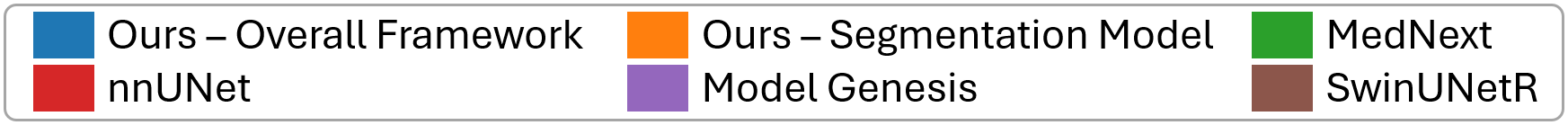}}
\end{minipage}

\vspace{-3mm}
\caption{Distribution of segmentation performance by subject across multiple Dice score thresholds. The upper panel (a higher count is better) shows the performance at higher Dice score thresholds ($\geq$ 0.5), indicating the number of subjects for each model achieving a better segmentation score than the threshold at the top of the graph. The lower panel (a lower count is better) shows the performance at lower Dice coefficient thresholds ($<$ 0.5), indicating the number of subjects for each model achieving a lower segmentation score than the threshold at the top of the graph. For the upper panel, higher counts are better while for the lower panel, lower counts are better. This provides insight into the reliability and consistency of each model in clinical settings. Higher bars at the upper panel suggest superior segmentation capabilities, while lower bars at the lower panel suggest reduced failure rates.}
\vspace{-2mm}
\label{fig:bar_suc_fail_subj}
\end{figure*}

This consistency is demonstrated in Fig. \ref{fig:box_subject} and Fig. \ref{fig:bar_suc_fail_subj}. In Fig. \ref{fig:box_subject}, the distribution of the Dice score across subjects using box plots is shown. The interquartile region of the proposed model and framework is consistently lower demonstrating a reduction in variability and increased consistency. Furthermore, the proposed approach reduces the number of subjects with low quality segmentation by 50\%, 28.5\%, 33.3\%, 28.5\%, and 12.5\% for low quality segmentation with Dice score thresholds of 0.1, 0.2, 0.3, 0.4, and 0.5 respectively as shown in Fig. \ref{fig:bar_suc_fail_subj}. On the other hand, the proposed approach increases the number of subjects with high quality segmentation by 1.4\%,  3.1\%, and 1.7\% for high quality segmentation with Dice score thresholds of 0.5, 0.6, and 0.7.

\subsubsection{Detection and Localization By Lesion}
\label{subsub-sec:lesion_det_loc}
\begin{figure}[htb!]
\vspace{-1mm}
    \centering
    \hspace{-5mm}\includegraphics[width = 0.82\linewidth, height=5cm, trim = {0.21cm 0.2cm 0.1cm 0.2cm}, clip]{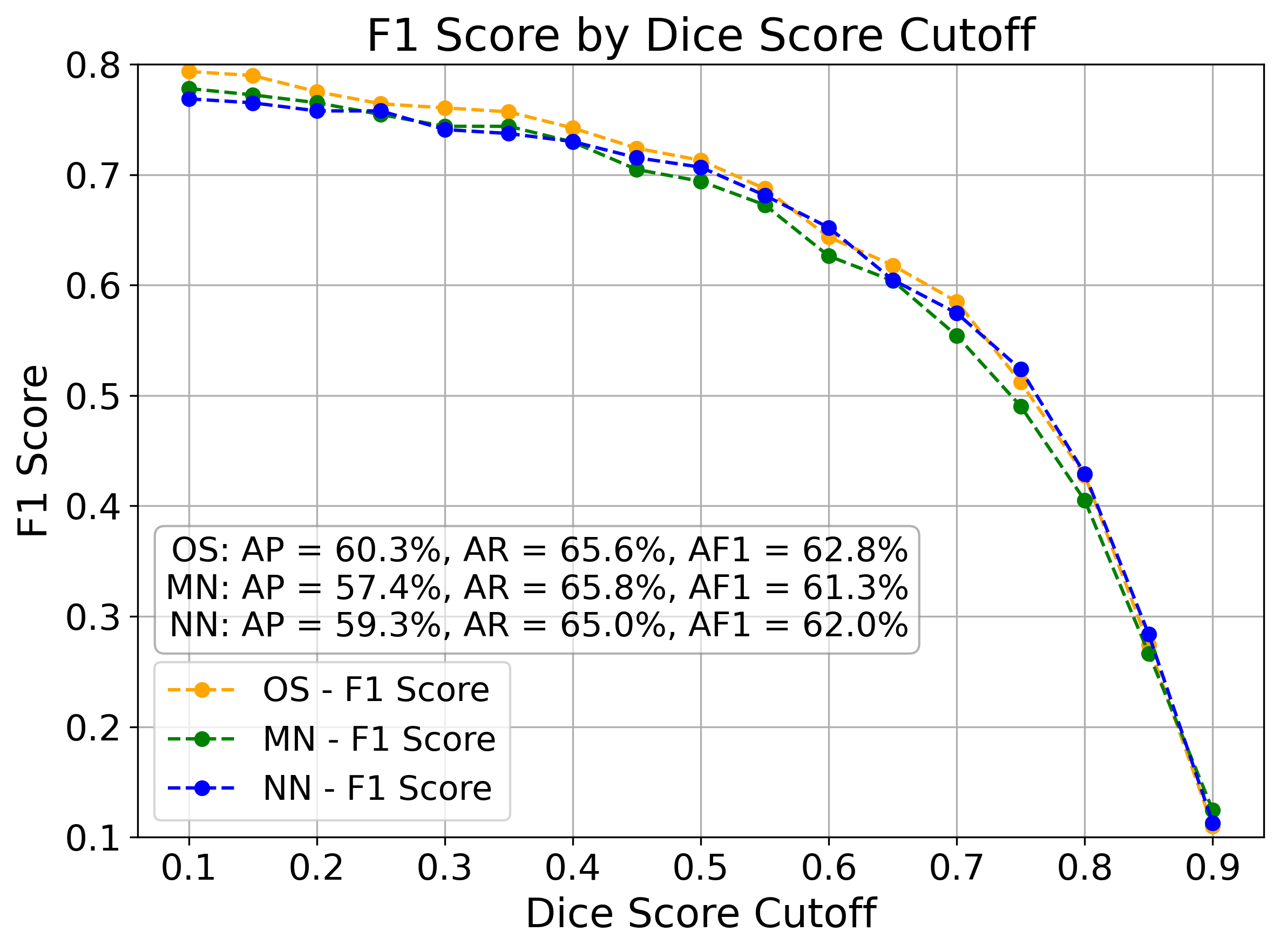}

\vspace{-3mm}
\caption{The lesion detection F1 score by Dice score cutoff. The average precision (AP), recall (AR), and F1 scores (AF1) across the Dice score cutoff range of 0.1 to 0.9 are outlined on the plot. Model names keys: OS = Ours, MN = MedNext, and NN = nnUNet.}
\vspace{-2mm}
\label{fig:res_by_lesion}
\end{figure}

To evaluate the proposed approach lesion detection performance,  we computed the precision, recall, and F1 score at different Dice score thresholds (0.1 to 0.9) and calculated the average precision, recall, and F1 scores across all the thresholds as shown in Fig. \ref{fig:res_by_lesion}. Overall, the average F1 score by lesion of the proposed approach is 62.8\% versus 61.3\% and 62.0\% for the MedNext and nnUNet models. In terms of localization, we used the average Euclidean distance between the center of ground truth and predicted segmentation by lesion. The difference in localization performance of our proposed approach with respect to the best performing benchmark model was relatively small with an average localization error for detected lesions of 3.44 mm versus 3.50 mm and a standard deviation of 5.40 mm versus 5.45 mm.

\subsection{Limitations and Future Prospective}
\label{sub-sec:lims}
The proposed framework is designed to improve the recovery and segmentation of lesions from multi-phase CT scans of the liver when compared to multi-channel segmentation models, which are the current state-of-the-art as shown in Table \ref{tab:results_main}. However, there are still instances (subjects) where models trained on individual phases outperform both the proposed framework and the multi-channel segmentation models. Therefore, a logical future extension of the framework is through the improvement of feature extraction from individual phase images, which we believe can be accomplished through either selective ensembling of models' predictions, or improved fusion of latent space representations from different phases at different stages of the model. Another challenge we observed in our proposed framework as well as other models is segmenting and recovering small lesions from the image volume, which remains an open and challenging problem for lesion segmentation.

\section{Conclusion}
\label{sec:conc}
We proposed a multi-stage segmentation framework for liver lesions in multi-phase CT scans. The proposed framework design enables an improved feature extraction from each of the phases when compared to the current state-of-the-art segmentation models. This enables the framework to improve the overall segmentation performance by 1.6\% while reducing performance variability across subjects by 8\%. As the backbone segmentation model of the framework, we proposed a UNet-like architecture that uses the ConvNext convolutional block in both the encoder and the decoder. In the skip connections we proposed the Coarse+Fine Feature Fusion \& Attention Module (C+F FFA) to enhance feature fusion and attention between the encoder and decoder, which improved segmentation accuracy by 0.4\% and reduced performance variability across subjects by 1.5\% when compared to the current-state-of-the-art model. Beyond liver lesion segmentation in multi-phase CT scans, we also tested the framework performance on segmenting brain tumors from multi-contrast MRI scans to further validate its ability to improve overall segmentation accuracy and reduce performance variability across subjects. 

\section*{Acknowledgments}
This work was supported by a grant from VinBrain, JSC.

%%%%%%%%%%%%%%%%%%%%%%%%%%%%%%%%%%%%%%%%%%%%%%%%%%%%%%%%%%%%%
%%Harvard
\bibliographystyle{model2-names.bst}
\biboptions{authoryear}
\bibliography{refs}

\end{document}